\documentclass[prd,11pt,a4paper, nofootinbib, preprintnumbers
,showpacs 
]{revtex4}
\usepackage{graphicx,epsf,epsfig}
\usepackage{amssymb}
\def\eq#1{{eq.~(\ref{#1})}}
\def\eqs#1#2{{eqs.~(\ref{#1})--(\ref{#2})}}

\def\vev#1{\left\langle #1\right\rangle}

\def\hbar{\hspace{0pt}\raisebox{1pt}{$-$} \hspace{-7pt} h}

\def\lowpos#1{\begin{array}{c}\mbox{}\\ \mbox{}\\ \mbox{}\\ #1 \end{array}}
\newcommand{\be}{\begin{equation}}
\newcommand{\ee}{\end{equation}}
\newcommand{\bd}{\begin{displaymath}}
\newcommand{\ed}{\end{displaymath}}
\newcommand{\bea}{\begin{eqnarray}}
\newcommand{\eea}{\end{eqnarray}}

\def\so10{$SO(10)$}

\begin{document}
%----------------------------------------------------------------------------------
\title{Towards a Complete Theory of Fermion Masses and Mixings \\ with $SO(3)$
Family Symmetry and 5d $SO(10)$ Unification}
%----------------------------------------------------------------------------------
\date{August 2, 2006}
\author{Stephen F. King}
\email{sfk@hep.phys.soton.ac.uk} \affiliation{School of Physics
and Astronomy, University of Southampton, SO16 1BJ Southampton,
United Kingdom}
\author{Michal Malinsk\'{y}}
\email{malinsky@phys.soton.ac.uk} \affiliation{School of Physics
and Astronomy, University of Southampton, SO16 1BJ Southampton,
United Kingdom}
%----------------------------------------------------------------------------------
%----------------------------------------------------------------------------------
%----------------------------------------------------------------------------------
\begin{abstract}
We construct a complete 4d model of fermion masses and mixings in the
Pati-Salam $SU(4)\otimes SU(2)_{L}\otimes SU(2)_{R}$ framework
governed by an $SO(3)$ gauged Family Symmetry.
The relevant low energy
effective Yukawa operators are constructed so that the $SO(3)$ flavons
enter at the simplest possible one-flavon level, with couplings
enforced by an additional $U(1)\times Z_2$ symmetry.
The simplicity of the flavon sector
allows the messenger sector to be fully specified,
allowing the ultraviolet completion of the model
at the 4d renormalizable level.
The model predicts approximate tri-bimaximal
lepton mixing via the
see-saw mechanism with sequential dominance, and vacuum alignment
of flavons, with calculable deviations described by the neutrino
sum rule. We perform a numerical analysis of the emerging charged fermion spectra and mixings.
The 4d model is shown to result from a 5d
orbifold GUT model based on $SO(3)\times SO(10)$, where small flavon
vacuum expectation values (VEVs)
originate from bulk volume suppression.
\end{abstract}
\maketitle
%%%%%%%%%%%%%%%%%%%%%%%%%%%%%%%%%%%%%%%%%%%
\section{Introduction}
%%%%%%%%%%%%%%%%%%%%%%%%%%%%%%%%%%%%%%%%%%%

The pattern of charged fermion masses and quark mixing angles
is described by 13 parameters in the minimal Standard Model
(3 charged lepton masses, 6 quark masses, 3 quark mixing angles and
1 quark CP violating phase).
The discovery of neutrino masses and lepton mixing angles
requires a further 9 parameters if neutrinos are
Majorana (3 neutrino masses, 3 lepton mixing angles and 3 lepton CP violating
phases), or 7 parameters if neutrinos are Dirac (2 fewer lepton phases).
The discovery of neutrino mass has consequently increased the number
of parameters in the flavour sector dramatically, providing further
motivation to understand the pattern of fermion masses and mixings.
The neutrino sector also provides additional data and clues 
which could enable this goal to be achieved.
For example, the neutrino sector is distinguished by having
two large mixing angles, one small mixing angle,
and very small mass eigenvalues \cite{Strumia:2006db}. 
Recently, as neutrino data has become more accurate,
intriguing patterns of mixing angles
in the lepton sector have begun to emerge \cite{Mohapatra:2006gs}.

A particularly promising strategy, leading to an effective reduction of
the number of independent Yukawa couplings, is the idea of Grand
Unified Theories (GUTs)
\cite{Pati:1974yy,Georgi:1974sy,Langacker:1980js}.
GUTs allow one to
understand the variety of Standard Model (SM) fermions as arising from
a few fundamental representations and often lead to strong
correlations amongst the corresponding low energy Yukawa couplings.  A
typical example could be the relation $m_{b}\approx  m_{\tau}$
\cite{Georgi:1974sy} emerging in a natural way in the class of GUT
models based on $SU(5)$ gauge symmetry \cite{Georgi:1974sy}.
Another example would be the
interesting correlation between $b-\tau$ unification and the large
atmospheric mixing angle \cite{Bajc:2002iw,Goh:2003sy} revealed in a
class of renormalizable $SO(10)$ models with type-II contribution
dominating the seesaw formula
\cite{Minkowski:1977sc,Gell-Mann:1980vs,Yanagida:1980xy,Glashow:1979nm,Mohapatra:1979ia,Magg:1980ut,Lazarides:1980nt,Mohapatra:1980yp}.  Although the SM
flavor problem is just one of many questions confronting GUTs (and
typically is not the main motivation to go beyond the SM gauge
structure) it is still very interesting that the detailed information
on the quark and lepton masses and mixing patterns can lead to severe
constraints on model building. For example it has been shown that
minimal renormalizable SUSY $SO(10)$ does not provide a consistent
description of the recent fermion mass and mixing data
\cite{Bajc:2002bv,Bajc:2002pg,Goh:2003hf,Bajc:2004fj,Bertolini:2004eq,Dutta:2004wv,Dutta:2004hp,Bertolini:2005qb,Bertolini:2006pe}.
In general GUT models do not provide a full understanding of the
observed pattern of quark and lepton masses and mixings. In
particular, without other assumptions there is usually no explanation
of the observed quark and lepton mass hierarchies, spanning the three
families.  Although the question of the fermion mass hierarchy has
been qualitatively addressed in approaches based on higher dimensional
orbifold GUTs (c.f. for instance
\cite{Altarelli:2001qj,Hall:2001rz,Haba:2002vc,Feruglio:2004zf,Dermisek:2001hp,Dienes:1998vg,Dienes:1998vh}
and references therein) in which the smallness of the first and second
family masses are traced back to their location in the extra
dimensions, see for instance
\cite{Haba:2002ek,Hall:2002ci,Haba:2002ve,Haba:2002vc,Kim:2002im,Kim:2004vk,Alciati:2006sw},
these constructions do not provide any understanding of the origin of
the small Cabibbo-Kobayashi-Maskawa (CKM) quark sector mixing together
with the observed bi-large mixing in the lepton sector.  Some GUT
models also include extra (global or local) Family Symmetry,
spontanously broken by flavon VEVs, in order to provide a more
predictive framework \cite{King:2003jb}.  The extra Family Symmetry acting nontrivially
among different flavors across the SM matter families provides
additional constraints on the Yukawa textures.  The problem of quark
and lepton mixing patterns has been recently addressed in models based
on extra continuous Family Symmetries like $U(1)$
\cite{Bando:2001bj,King:2000ge,Ling:2002nj,Ling:2003kr,Kane:2005va,Grimus:2002zh},
$SU(2)$ \cite{Blazek:1999hz,Raby:2003ay}, $SU(3)$
\cite{King:2001uz,King:2003rf}, $SO(3)$ \cite{Barbieri:1999km,Antusch:2004xd} or discrete subgroups of continuos
symmetries like for instance $D_{5}$ \cite{Hagedorn:2006ir}, $D_{4}$
\cite{Grimus:2004rj}, $S_{4}$ \cite{Hagedorn:2006ug}, $A_{4}$
\cite{Babu:2002dz}.

One of the challenges facing such GUT and Family Symmetry models
is to provide a convincing 
explanation of the observed (approximate) tri-bimaximal lepton mixing,
corresponding to a maximal atmospheric angle
$\tan \theta_{23}\approx 1$, a zero reactor angle $\theta_{13}\approx
0$ and a solar angle
$\sin \theta_{12}\approx 1/\sqrt{3}$ \cite{Harrison:2002er}.
The possibility that the tri-bimaximal neutrino
mixing matrix involves
square roots of simple ratios motivates models in which the mixing
angles are independent of the mass eigenvalues. One such class of
models are see-saw models with sequential dominance of
right-handed neutrinos 
\cite{King:1998jw,King:1999cm,King:1999mb,King:2002nf}.
In sequential dominance, a neutrino mass hierarchy is shown to
result from
having one of the right-handed neutrinos give the dominant
contribution to the
see-saw mechanism, while a second right-handed neutrino gives the
leading sub-dominant contribution, leading to a neutrino mass
matrix
with naturally small determinant.
In a basis where the right-handed neutrino mass matrix
and the charged lepton mass matrix are
diagonal,
the atmospheric and solar neutrino mixing angles are determined in
terms of ratios
of Yukawa couplings involving the dominant and subdominant
right-handed neutrinos, respectively. 
If these Yukawa couplings
are simply related in some way, then it is possible for simple
neutrino mixing angle relations, such as appear in tri-bimaximal
neutrino mixing, to emerge in a simple and natural way,
independently of the neutrino mass eigenvalues.
Specifically,
if the dominant right-handed neutrino couples
equally to the second and third family,
with a zero coupling to the first family, then this will
result in a maximal atmospheric mixing angle
$\tan \theta_{23}\approx 1$.
If the leading subdominant right-handed neutrino couples
equally to all three families, and if these couplings are
orthogonal to the couplings of the dominant right-handed neutrino,
then this will result in a tri-bimaximal solar neutrino mixing angle
$\sin \theta_{12}\approx 1/\sqrt{3}$, and a zero reactor angle
$\theta_{13}\approx 0$, assuming that the third right-handed neutrino
is completely decoupled from the see-saw mechanism.
This is called constrained sequential dominance (CSD) \cite{King:2005bj}.
In realistic models there will be 
corrections to tri-bimaximal mixing from charged lepton mixing,
resulting in testable predictions and sum rules for lepton mixing
angles \cite{King:2005bj,Antusch:2005kw}.

In order to achieve the CSD Yukawa relations 
it seems to be necessary to introduce a non-Abelian horizontal Family Symmetry
spanning the three families.
The Family Symmetry is then broken by flavons, 
and effective Yukawa operators may
be constructed where the aligned flavon VEVs 
provide the required CSD relations
between the Yukawa couplings. 
This strategy has been followed for models based on the 
Family
Symmetry 
$SO(3)$ \cite{King:2005bj}, $SU(3)$ \cite{deMedeirosVarzielas:2005ax}, 
or their discrete subgroups 
\cite{deMedeirosVarzielas:2005qg,deMedeirosVarzielas:2006fc}
(see also \cite{Altarelli:2005yx}). 
The choice of $SU(3)$, or a discrete subgroup
of it such as $\Delta (27)$ \cite{deMedeirosVarzielas:2006fc}, 
has the advantage that it enables both the
left and right handed chiral fermions to both transform under the
Family Symmetry as triplets, permitting
unification into a single $SU(3)\times SO(10)$ multiplet
$(3,16)$. The lowest order effective Yukawa operators for $SU(3)$ must then
involve a minimum of two anti-triplet flavon insertions.
The choice of $SO(3)$, or a discrete subgroup of it such as
$A_4$ \cite{deMedeirosVarzielas:2005qg}, 
requires that only one type of chiral fermion transform
under the Family Symmetry, while the other type is a family singlet,
in order to avoid trivial Family Symmetry contractions.
The disadvantage is that it seems to not allow a similar
unification into a single $SO(3)\times SO(10)$ multiplet
$(3,16)$, however it has the advantage that the
lowest order effective Yukawa operators for $SO(3)$
involve only one triplet flavon insertion, and are therefore simpler.
However, in practice, this advantage has so far not been exploited
since the lowest order effective operator has only been used for
the largest Yukawa coupling associated with the third family,
while the operators associated with the first and second family
were assumed to involve three flavons, in order to account for the
required suppression, where the resulting scheme required
large additional symmetries \cite{King:2005bj}.

In the present paper we shall exploit the simplicity of
$SO(3)$ by assigning to all three families effective Yukawa
operators involving just single flavon insertions
in the Dirac sector.
We assume that there are essentially two types of flavons,
one type which develops large VEV and may be used to describe the
third family, and a second type which develops a small VEV and
will be assigned to
the lighter families. In our approach the main role of the
horizontal symmetry (realized at the quantum
level in terms of extra interactions of matter with relevant
flavon fields) is to explain the {\it correlations} among the Yukawa
entries rather than their {\it hierarchy}, which is accounted for
by the small flavon VEVs.
Since the flavons enter the
effective operators at the lowest possible level, this allows us to
reduce the usually cumbersome extra symmetries considerably, as
there is no need to suppress wide classes of effective operators
up to high order in the number of flavon insertions.
The resulting relative simplicity of the flavon sector encourages us to
go beyond the effective non-renormalizable
Yukawa operator description in 
\cite{King:2005bj,deMedeirosVarzielas:2005ax,deMedeirosVarzielas:2005qg,deMedeirosVarzielas:2006fc}
and construct explicitly the renormalizable messenger
sector, leading to an ultraviolet completion of the model.
The messenger sector allows for effectively different
expansion parameters in the different charged sectors,
and the effective Yukawa and Majorana matrices are then constructed.
We perform a numerical analysis which 
shows that the model provides an excellent
fit to the charged fermion mass spectrum.
The model also predicts approximate tri-bimaximal
lepton mixing via CSD due to vacuum alignment
of flavon VEVs, with calculable deviations described by the neutrino
sum rule. The strong hierarchy in the charged
fermion sector gets cancelled in the neutrino sector,
via the see-saw mechanism with sequential dominance,
leading to $m_2 \sim m_3$ for the lowest order neutrino masses, 
with the mild neutrino hierarchy $m_2 /m_3 \sim 1/5$  
produced by higher order corrections 
necessarily present in the model.
Finally we show that the 4d model here can result from a 5d
orbifold GUT model based on $SO(3)\times SO(10)$,
leading to a full $SO(10)$ unification of the $SO(3)$ model, and
an explanation of 
the small flavon VEV responsible for the fermion mass hierarchy
in terms of bulk volume suppression.
The synthesis of non-Abelian Family Symmetry
with orbifold GUTs provides an attractive way of
simplifying the Yukawa operators required
by explaining the fermion mass hierarchy in terms of  
a single suppressed flavon VEV rather than a 
higher order operator. Such a simplification of the Yukawa
operators at the non-renormalizable level is instrumental
in allowing us to provide
the ultraviolet completion of the model in terms of an
explicit messenger sector.

The layout of the remainder of the paper is as follows.
In section II we discuss the 4d model at the effective operator level,
specifying the symmetry and field content of the model,
and performing a full operator analysis
of the effective Dirac  and Majorana operators.
We show that it leads to approximate tri-bimaximal lepton mixing
and a normal neutrino mass hierarchy.
In section III we discuss the complete 4d model including the
messenger sector responsible for the effective operators.
In section IV we perform a numerical analysis of the model,
where we show that the parameters of the model can provide
a successful fit for the quark masses and mixings using the charged
lepton masses as inputs.
In section V we discuss the embedding of the model into a 5d
$SO(3)\times SO(10)$ orbifold GUT model, in which small flavon
VEVs can be accounted for by volume suppression, and
the full $SO(10)$ unification of the model is manifested.
Section VI concludes the paper.

%==========================
\section{The 4d effective non-renormalizable model}
%==========================
 \label{4dsection}
%==========================
\subsection{The symmetry}
%==========================

We work in a class of supersymmetric Pati-Salam models based on
the gauge symmetry group $SU(4)\otimes SU(2)_{L}\otimes
SU(2)_{R}$, which is supposed to be spontaneously broken at some
high scale $M_{G}$ (typically $\gtrsim 10^{15}$ GeV) to the
ordinary $SU(3)_{c}\otimes SU(2)_{L}\otimes U(1)_{Y}$ of the
minimal supersymmetric standard model (MSSM). Though this is not a
fully unified description of both the strong and electroweak
interactions (just as the Standard Model does not fully
unify the electromagnetic and weak nuclear force), this structure
is liberal enough to let the left-handed matter fields transform
nontrivially under the horizontal $SO(3)$ unlike the right-handed
SM fermions (that are $SO(3)$ singlets\footnote{This choice leads
to the typical correlations among entries in columns of the
relevant Dirac Yukawa matrices (in LR notation), just in a way it
is expected in a class of lepton sector models with sequential
dominance.}), while still rigid enough to give rise to a set of
nontrivial correlations among the quark and lepton Yukawa
couplings. The horizontal $SO(3)$ is a gauged Family Symmetry
while the extra $U(1)\otimes Z_{2}$
factors are supposed to be  approximate\footnote{This is there
namely to avoid problems with Goldstone bosons and/or topological
defects below the extra symmetry breakdown scale.} global
symmetries of the model at the Pati-Salam level. We expect these
symmetries to be broken spontaneously by the VEVs of a set of
flavon fields transforming trivially under the gauge symmetry\footnote{Thus 
keeping the GUT-like gauge coupling convergence
intact. (However, there is no need to demand an exact $SU(5)$ or
$SO(10)$-like gauge coupling unification here as $SU(4)\otimes
SU(2)_{L}\otimes SU(2)_{R}$ is not a simple group.)}.

%==========================
\subsection{The field content}
%==========================

We assume the minimal Pati-Salam matter content and let
the left-handed matter fermions (transforming like $(4,2,1)$ under
$SU(4)\otimes SU(2)_{L}\otimes SU(2)_{R}$) denoted by $\vec{F}$
form a triplet under the $SO(3)$ flavor symmetry, while the
right-handed components $F_{1}^{c}$, $F_{2}^{c}$ and $F_{3}^{c}$
(behaving like $(\overline{4},1,2)$ under the PS symmetry) are
supposed to be $SO(3)$ singlets. There is one copy of the Higgs
bidoublet driving the SM spontaneous symmetry breakdown and a pair
of  Higgs fields (denoted by $H\oplus \overline{H}$ and $H'\oplus
\overline{H'}$) responsible for the proper breaking of the PS
symmetry and the Majorana masses of neutrinos. Last, there is an
extra Higgs field $\Sigma$ transforming as $(15,1,3)$ of
Pati-Salam symmetry that gives rise to the desired Georgi-Jarlskog
\cite{Georgi:1979df} Clebsch factors in the charged matter sector
while keeping the effective neutrino Dirac Yukawa couplings intact.

Concerning the flavon sector, we follow the generic construction
in \cite{King:2005bj}. Since the model will involve a
minimal number of flavon insertions, their extra charges are
chosen to be opposite to those of $F_{1}^{c}$, $F_{2}^{c}$
and $F_{3}^{c}$ so that a particular flavon is associated with
a particular column of the Yukawa matrix, at lowest order.
The full set of the effective theory matter, Higgs and flavon fields and their transformation properties are given in Table \ref{tab-fields}.
\begin{table}[ht]
\centering
\begin{tabular}{|c|c|c|c|c|}
\hline
field & $SU(4)\otimes SU(2)_{L}\otimes SU(2)_{R}$ & $SO(3)$ & $U(1)$  & $Z_{2}$\\
\hline
$\vec{F}$ &  $(4,2,1)$ & $3$ & 0 & $+$\\
$F^{c}_{1}$   & $(\overline{4},1,2)$ & $1$ & $+2$ & $-$ \\
$F^{c}_{2}$  & $(\overline{4},1,2)$ & $1$ & $+1$ & $+$\\
$F^{c}_{3}$   & $(\overline{4},1,2)$ & $1$ & $-3$ & $-$\\
\hline
$h$  & $(1,2,2)$  &$1$ & 0 & $+$\\
$H$, $\overline{H}$  & $(4,1,2)$, $(\overline{4},1,2)$ & $1$ &
$\pm 3$ & $+$
\\
$H'$, $\overline{H'}$   & $(\overline{4},1,2)$,
$(\overline{4},2,1)$ & $1$ & $\mp 3$ & $+$
\\
\hline
$\Sigma$  & $(15,1,3)$  &$ 1 $ & -1 & $-$\\
\hline
$\vec{\phi}_{3}$  & $(1,1,1)$ & $3$ & $+3$ & $-$ \\
$\vec{\phi}_{23}$  & $(1,1,1)$ & $3$ & $-2$ & $-$\\
$\vec{\phi}_{123}$  & $(1,1,1)$ & $3$ & $-1$ & $+$\\
$\vec{\phi}_{12}$  & $(1,1,1)$ & $3$ & $0$ & $+$\\
$\vec{\tilde{\phi}}_{23}$  & $(1,1,1)$ & $3$ & $0$ & $-$ \\
\hline
\end{tabular}
\caption{\label{tab-fields}  The basic Higgs, matter and flavon
content of the model. }
\end{table}

With all this information at hand, one can easily
infer the shape of the lowest level effective operators allowed by
the gauge and extra symmetries. Let us start with the Dirac Yukawa
texture, that exhibit the effects of the $SO(3)$ horizontal
symmetry in its full glory, in particular in the neutrino sector.

%==========================
\subsection{The Dirac sector}
%==========================
The main feature of our construction is the simplicity of the
leading effective operators (responsible in particular for the
desired tri-bimaximal structure of the neutrino Dirac and Majorana Yukawa matrices, c.f.
section \ref{Majorana}), in particular the fact that they all emerge at one
flavon insertion level, with all the advantages over the former
constructions (the simplicity of the extra symmetries and
the would-be Froggatt-Nielsen messenger sector):
%----
\be\label{diracsuperpot1} W_{Y}^{leading}=\frac{1}{M}
y_{23}\vec{F}.\vec{\phi}_{23}F_{1}^{c}h+ \frac{1}{M}
y_{123}\vec{F}.\vec{\phi}_{123}F_{2}^{c}h+ \frac{1}{M}
y_{3}\vec{F}.\vec{\phi}_{3}F_{3}^{c}h +\ldots \ee where $M$ stands
for the masses of the relevant Froggatt-Nielsen messenger fields
while the elipses cover the subleading terms necessary for the
proper desription of the quark sector details (hierarchies and the
CKM mixing parameters, the first generation masses etc.): 
\be\label{diracsuperpot2}
W_{Y}^{subl.}=\frac{1}{M^{2}}
y_{GJ}\vec{F}.\vec{\tilde{\phi}}_{23}F_{2}^{c}\Sigma h+
\frac{1}{M^{2}} y_{12}\vec{F}. (\vec{\phi}_{3}\times
\vec{{\phi}}_{12}) F_{3}^{c}h+ \frac{1}{M^{3}}
\tilde{y}_{23}\vec{F}.\vec{\tilde{\phi}}_{23}(\vec{\phi}_{3}.\vec{\tilde{\phi}}_{23})
F_{3}^{c}h +\ldots \ee
As we claimed before, we exploit the $SO(3)$ horizontal symmetry
to understand the correlations among the various Yukawa entries
rather than their exact hierarchy\footnote{As a matter of fact,
such attempts are questionable indeed, because to understand
hierarchies an extra symmetry does not help much unless a scale at
which it becomes broken is specified, i.e. there is always an
extra ingredient needed to accomplish such a goal.}. Instead, we
equip some of the flavon and Higgs VEVs with extra suppression
factors (with respect to their natural values dictated by the
relevant symmetry breaking scales) and let these factors tell
their favorite values just upon fitting all the quark and lepton
data. Remarkably enough, there is an option to connect all
of them to just one universal suppression scale, that could find a natural
justification for instance in higherdimensional constructions.

As an example, consider the CSD structure of the neutrino Dirac
Yukawa matrix $Y^{\nu}$ emerging from \eqs{diracsuperpot1}{diracsuperpot2} (for further details see also formula (\ref{Ynu0})). 
The lepton mixing data do
not specify the overall magnitudes of its first and second
columns, only the correlations among their entries. This is
precisely where the horizontal symmetry is supposed to play an
important role. Only after its embedding into a (partially)
unified framework like the Pati-Salam gauge model, their
correlations with the quark sector Dirac Yukawas trigger the need
of a particular suppression of the first and second column entries
with respect to the third column ones.

Thus, it is quite natural to let these two requirements of
intrinsically different origins be justified from two different
sources like we propose here -- the $SO(3)$ symmetry shall govern
the rescaling-invariant quantities like the lepton mixing angles
(in seesaw-type models) while the charged matter sector
hierarchies emerge from the suppression of the flavon VEVs $\vev{\phi_{23}}$, $\vev{\phi_{123}}$ and $\vev{\phi_{12}}$ (and similarly for $H'$ Higgs field). We let these VEVs (driven by
naturalness to roughly the same order of magnitude corresponding
to the scale of the $SO(3)$ (and Pati-Salam) symmetry breaking, at least if it is
one-step) be suppressed by extra factors called $\delta_{23}$, $\delta_{123}$ and
$\delta_{12}$ with respect to the VEV of $\phi_{3}$, namely:
\be\label{flavonhiearchy} |\vev{\phi_{123}}|\sim
\delta_{123}|\vev{\phi_{3}}|, \qquad |\vev{\phi_{23}}|\sim
\delta_{23} |\vev{\phi_{3}}| \quad {\rm and } \quad
|\vev{\phi_{12}}|\sim \delta_{12} |\vev{\phi_{3}}| 
\ee
As we shall see later in section \ref{5dmodel} all these suppression factors can be justified in terms of a universal suppression factor $\delta$ coming from extra dimensional dynamics. 
Note that we keep the VEV of the ``Georgi-Jarlskog'' flavon $\tilde\phi_{23}$ at the natural scale $|\vev{\tilde\phi_{23}}|\sim |\vev{\phi_{3}}|$ and let the slight suppression of the second generation masses come from the higher level nature of the relevant effective operator in \eq{diracsuperpot2}. 

In a
similar manner, the Majorana sector structure shall be affected by
the requirement of having the VEV of $H'$ well below the VEV of
$H$, namely \be\label{Higgshierarchy} |\vev{H'}|\sim
\delta_{H}|\vev{H}|. \ee
The alignment of the flavon VEVs will be assumed to be given by
\cite{King:2005bj}:
\be
\label{123}
\langle\phi_{3}\rangle^{T}\propto (0,0,1), \ \
\langle\phi_{23}\rangle^{T}\propto (0,1,-1), \ \
\langle\tilde{\phi}_{23}\rangle^{T}\propto (0,1,1), \ \ 
\langle\phi_{123}\rangle^{T}\propto (1,1,1), \ \ 
\langle\phi_{12}\rangle^{T}\propto (1,1,0). 
\ee
and will not be discussed further here.

After the appropriate flavor symmetry
breaking 
% driven by $\phi_{3}$ and $\phi_{23}+\phi_{123}$ flavons
this structure gives rise to the Dirac mass matrices like
(in an obvious symbolic LR notation):
\be
\label{diracpart}
Y^{f}_{LR}=\left(\begin{array}{ccc}
0 &y_{123} \varepsilon^f_{123} &  {y}_{12} \varepsilon^f_{12} \varepsilon^f_{3} \\
y_{23} \varepsilon^f_{23} & y_{123} \varepsilon^f_{123}+  C^f y_{GJ}\tilde{\varepsilon}^f_{23}\sigma & \tilde{y}_{23}(\tilde{\varepsilon}_{23})^{2} \varepsilon_{3} \\
-y_{23}\varepsilon^f_{23} &y_{123} \varepsilon^f_{123} + C^f
y_{GJ}\tilde{\varepsilon}^f_{23}\sigma & y_{3}\varepsilon^f_{3}
\end{array}\right)
\ee
where $C^f=-2,0,1,3$ for $f=u,\nu, d,e$ are the
traditional Clebsch-Gordon coefficients responsible for the
distinct charged sector hierarchies, $\sigma$ denotes the
(normalized) VEV of the Georgi-Jarlskog field $\sigma\equiv
\langle\Sigma\rangle/M_f$ and $\varepsilon^f_{x}$ stands for the
various flavon VEV factors $\langle\phi_x\rangle/M_f$.

Note that the choice of the $\Sigma$ field giving rise to the
desired Georgi-Jarlskog Clebsch factor in the charged sector
Yukawas $Y^{e,u,d}$ is practically unique. Since we need to
preserve the tight CSD homogeneity of the second column of the
neutrino dirac mass matrix, the effect of the $\Sigma$ VEV should
be strongly suppressed in $Y^\nu$ by the relevant Clebsch factor
of $\Sigma$ VEV in the neutrino direction. Then, $\Sigma$
transforming like $(15,1,3)$ under $SU(4)_{C}\otimes SU(2)_{R}\otimes SU(2)_{L}$ is the simplest choice that can satisfy this requirement.

%------------------------------------------
\subsection{The Majorana neutrino sector \label{Majorana}}
%------------------------------------------

Assuming the vacuum alignment as in the previous subsection,
the neutrino Yukawa matrix takes the form:
\be
\label{Ynu0}
Y^{\nu}_{LR}= \left(\begin{array}{ccc}
0 &y_{123} \varepsilon^{\nu}_{123} &  {y}_{12} \varepsilon^{\nu}_{12} \varepsilon^{\nu}_{3}\\
y_{23} \varepsilon^{\nu}_{23}  & y_{123} \varepsilon^{\nu}_{123} & \tilde{y}_{23}(\tilde{\varepsilon}^{\nu}_{23})^{2} \varepsilon^{\nu}_{3}\\
-y_{23}\varepsilon^{\nu}_{23} & y_{123} \varepsilon^{\nu}_{123} &
y_{3}'\varepsilon^{l}_{3}+y_{3}\varepsilon^{\nu}_{3}
\end{array}\right)
\ee
Assuming that the right-handed neutrino associated with the
first column gives the dominant contribution to the see-saw mechanism,
the second right-handed neutrino gives the leading
subdominant contribution, and the third column gives the
smallest contribution, then this form of neutrino Yukawa matrix
corresponds to constrained sequential dominance (CSD),
and will lead to tri-bimaximal
lepton mixing as discussed in \cite{King:2005bj}.

The Majorana right-handed neutrino mass matrix must be approximately diagonal,
so as not to lead to significant corrections to the Yukawa
matrix in the diagonal right-handed neutrino mass basis,
and in addition it must be sufficiently hierarchical to ensure that
the right-handed neutrinos dominate sequentially as described
above. The (leading order) structure of the neutrino Majorana mass matrix
is triggered by the choice of the $U(1)\times Z_{2}$ charges of
the heavy Higgs fields $H$ and $H'$. Assuming the hierarchy among
the VEVs of the flavon and Higgs fields given by
\eqs{flavonhiearchy}{Higgshierarchy} the lowest level effective
operators allowed by the extra symmetries are: 
\be\label{majlead} W_{M}^{lead.}=
\frac{1}{M_{\nu}^{3}}w_{1}{F_{1}^{c}}^{2}H H' \phi_{23}^{2}+
\frac{1}{M_{\nu}^{3}}w_{2}{F_{2}^{c}}^{2}H H' \phi_{123}^{2}+
\frac{1}{M_{\nu}}w_{3}{F_{3}^{c}}^{2}H^{2}+\ldots 
\ee
Assuming the relevant messengers to be the same as
for the Dirac neutrino sector (see later)
these terms generate a diagonal Majorana
mass matrix
$$M^{\nu}_{RR}={\rm diag}(w_{1} {\varepsilon^{\nu}_{23}}^{2} \delta_{H},w_{2}
{\varepsilon^{\nu}_{123}}^{2} \delta_{H},w_{3})M_3
$$
where as before $\varepsilon^\nu_{x}$ denotes $\langle\phi_x\rangle/M$.
The see-saw formula $m_\nu = Y^{\nu}_{LR}{M^{\nu}_{RR}}^{-1}{Y^{\nu}_{LR}}^T v^{2}$
leads to three contributions to the light neutrino mass matrix, from
each of the three right-handed neutrinos,
the first and second of order $\delta_{H}^{-1}v^2/M_3$,
and the third of order ${\varepsilon^{l}_{3}}^2v^2/M_3$.
With sufficiently small $\delta_{H}$ the third right-handed
neutrino becomes decoupled and irrelevant for the see-saw mechanism.
Such a simple Majorana
structure has several noteworthy features. In particular, the
would-be $\delta$-suppressions associated to the $\phi_{23}$ and
$\phi_{123}$ VEVs entering the Dirac neutrino Yukawa through the
leading operators given in \eq{diracsuperpot1} is cancelled
in the seesaw formula by the suppression factors present in
$M^{\nu}_{RR}$. The $\varepsilon^{\nu}_{x}$ suppression factors similarly
cancel, leading to $m_2 \sim m_3$,
in contrast to the strong hierarchy in the charged matter spectra.
Note, however, that $m_{1}\ll m_2$. 

The above see-saw cancellations, though welcome from the point of view
of making the hierarchy between $m_2$ and $m_3$ mild,
are apparently too efficient and at leading order lead to
no hierarchy at all, $m_2 \sim m_3$. However,
apart from the leading terms given above, the extra symmetries allow
for many subleading terms, for instance \footnote{Here we
typically omit the allowed $SO(3)$ contractions that drop out in
the mass matrix because of the orthogonality of the relevant
flavon VEVs, for instance $\frac{1}{M^{5}}{F_{2}^{c}}^{2}H'H'
\vec\phi_{3}^{2}(\vec\phi_{23}.\vec{\tilde\phi}_{23}) $,
$\frac{1}{M^{3}}{F_{1}^{c}}{F_{2}^{c}}H'H'
\vec\phi_{3}.\vec\phi_{12}$ etc.}:
$$
W_{M}^{subl.}= \frac{w_{4}}{M^{4}}{F_{1}^{c}}^{2}H'H' (\vec\phi_{3}
\times \vec\phi_{123}).\vec{\tilde\phi}_{23}+
\frac{w_{5}}{M^{4}}{F_{2}^{c}}^{2}H H' (\vec\phi_{23} \times
\vec\phi_{12}).\vec{\tilde\phi}_{23}+
\frac{w_{6}}{M^{5}}{F_{2}^{c}}^{2}H'H' (\vec\phi_{3}. \vec\phi_{23})
(\vec\phi_{3}.\vec{\tilde\phi}_{23}) +
$$
$$
+
\frac{w_{7}}{M^{4}}{F_{1}^{c}}{F_{2}^{c}} H H' (\vec\phi_{23} \times
\vec\phi_{123}).\vec{\phi}_{12}+
\frac{w_{8}}{M^{5}}{F_{1}^{c}}{F_{2}^{c}} H'H' (\vec\phi_{3}
.\vec{\tilde\phi}_{23})(\vec\phi_{12}
.\vec{\tilde\phi}_{23})+\frac{w_{9}}{M^{4}}{F_{1}^{c}}{F_{3}^{c}} H H'
(\vec\phi_{3} \times \vec\phi_{23}).\vec{\phi}_{12}+ 
$$
\be\label{offdiagonal}\label{extraoperators1}\label{majsublead} 
+\frac{w_{10}}{M^{6}}{F_{2}^{c}}{F_{3}^{c}} H H
(\vec\phi_{123} \times \vec\phi_{23}).\vec{\tilde{\phi}}_{23} (\vec{\tilde{\phi}}_{23}.\vec\phi_{123})+\frac{w_{11}}{M^{6}}{F_{2}^{c}}{F_{3}^{c}} H H'
(\vec\phi_{3} \times \vec\phi_{23}).\vec{\tilde{\phi}}_{23} (\vec{{\phi}}_{23}.\vec\phi_{3})
+\ldots
\ee
With this information at hand the structure of the Majorana mass matrix reads\footnote{Only the leading contributions (in number of  suppressions in $\delta_{H}$ and $\varepsilon_{23}^{\nu}$, $\varepsilon_{123}^{\nu}$ and $\varepsilon_{12}^{\nu}$) are displayed.}
 \be\label{majoranapart2}
M^{\nu}_{RR}= \left(\begin{array}{ccc}
{\cal O}(\varepsilon^{\nu 2}_{23}\delta_{H},\varepsilon^{\nu}_{123}\tilde{\varepsilon}^{\nu}_{23}\delta_{H}^{2})& {\cal O}(\varepsilon^{\nu}_{123}\varepsilon^{\nu}_{23}\varepsilon^{\nu}_{12}\delta_{H},\varepsilon^{\nu}_{12}\varepsilon^{\nu}_{3}\tilde{\varepsilon}^{\nu 2}_{23}\delta_{H}^{2}) & {\cal O}(\varepsilon_{3}^{\nu}\varepsilon_{23}^{\nu}\varepsilon_{12}^{\nu}\delta_{H}) \\
. & {\cal O}(\varepsilon^{\nu 2}_{123}\delta_{H},\varepsilon^{\nu}_{12}\varepsilon^{\nu}_{23}\tilde{\varepsilon}^{\nu}_{23}\delta_{H},\ldots) & {\cal O}(\varepsilon^{\nu 2}_{123}\varepsilon^{\nu}_{23}\tilde{\varepsilon}^{\nu 2}_{23},\varepsilon^{\nu 2}_{3}\varepsilon^{\nu 2}_{23}\tilde{\varepsilon}^{\nu}_{23}\delta_{H}) \\
. & . &{\cal O}(1)
\end{array}\right)
\frac{\vev{H}^{2}}{M} \ee
Assuming $\delta_{123}\sim\delta_{23}\sim\delta_{12}\sim\delta_{H}\equiv \delta$ (leading to $\varepsilon^{\nu}_{123}\sim\varepsilon^{\nu}_{23}\sim\varepsilon^{\nu}_{12}\sim \delta \varepsilon_{3}^{\nu}\sim \delta \tilde{\varepsilon}_{23}^{\nu}$, c.f. formula (\ref{flavonhiearchy})) the lepton mixing angles emerging from the Majorana sector are:
\be\label{majoranaangles} \theta^{RR}_{12}\sim  {\cal
O}(\tilde{\varepsilon}^{\nu 2}_{23}), \quad
\theta^{RR}_{13}\sim {\cal O}( \delta^{3}\varepsilon_{3}^{\nu}), \quad
\theta^{RR}_{23}\sim {\cal O}(\delta^{3}\varepsilon_{3}^{\nu}\tilde{\varepsilon}_{23}^{\nu}) \ee
Since $\varepsilon_{3}^{\nu}\sim \tilde{\varepsilon}_{23}^{\nu}\ll 1$ (see section \ref{majoranamesssector})
we find that these angles are
small enough not to disturb the required CSD Dirac sector
correlations significantly, which would spoil the
tri-bimaximal prediction.
However the beneficial consequence of the effective operators is that
one has enough room to smear the unwanted degeneracy of the first
and second heavy Majorana masses restoring the validity of the
second CSD hierarchy condition \cite{King:2005bj}, leading to $m_2/m_3 \sim 1/5$
by a suitable choice of parameters.

%==========================
\section{The 4d renormalizable model}
%==========================

%------------------------------------------
\subsection{The messenger sector}
%------------------------------------------
We now present a renormalizable 4d theory which gives rise to the
effective non-renormalizable operators of the previous section.
The effective non-renormalizable operators will arise from
the exchange of heavy messenger fields.
In this subsection we shall describe the messenger sector
responsible for the effective Dirac operators.
Note that the construction of the full model at the
renormalizable level is greatly facilitated by the
simplicity of the model at the effective operator level,
in particular the fact that the simplest operators
correspond to the insertion of only one flavon.

At the level of one flavon insertion operators dominating the
Dirac Yukawa structures there  are in principle two distinct
classes of Froggatt-Nielsen operators behind, namely:
\be
\label{messengertypes}
\lowpos{\rm type\, 1:}
\parbox{5cm}{\includegraphics[width=5cm]{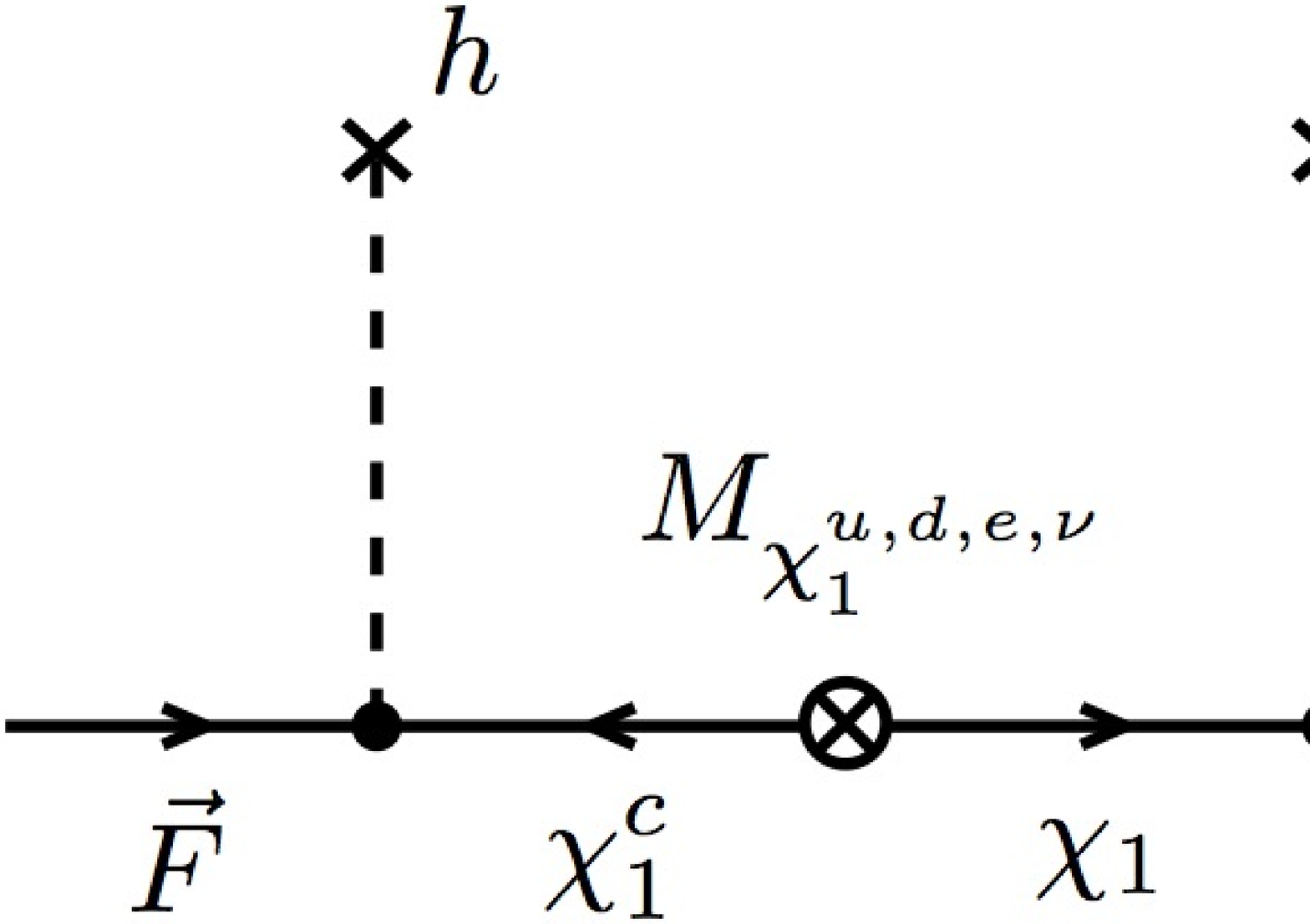}}\qquad
\lowpos{\rm type\, 2:}
\parbox{5cm}{\includegraphics[width=5cm]{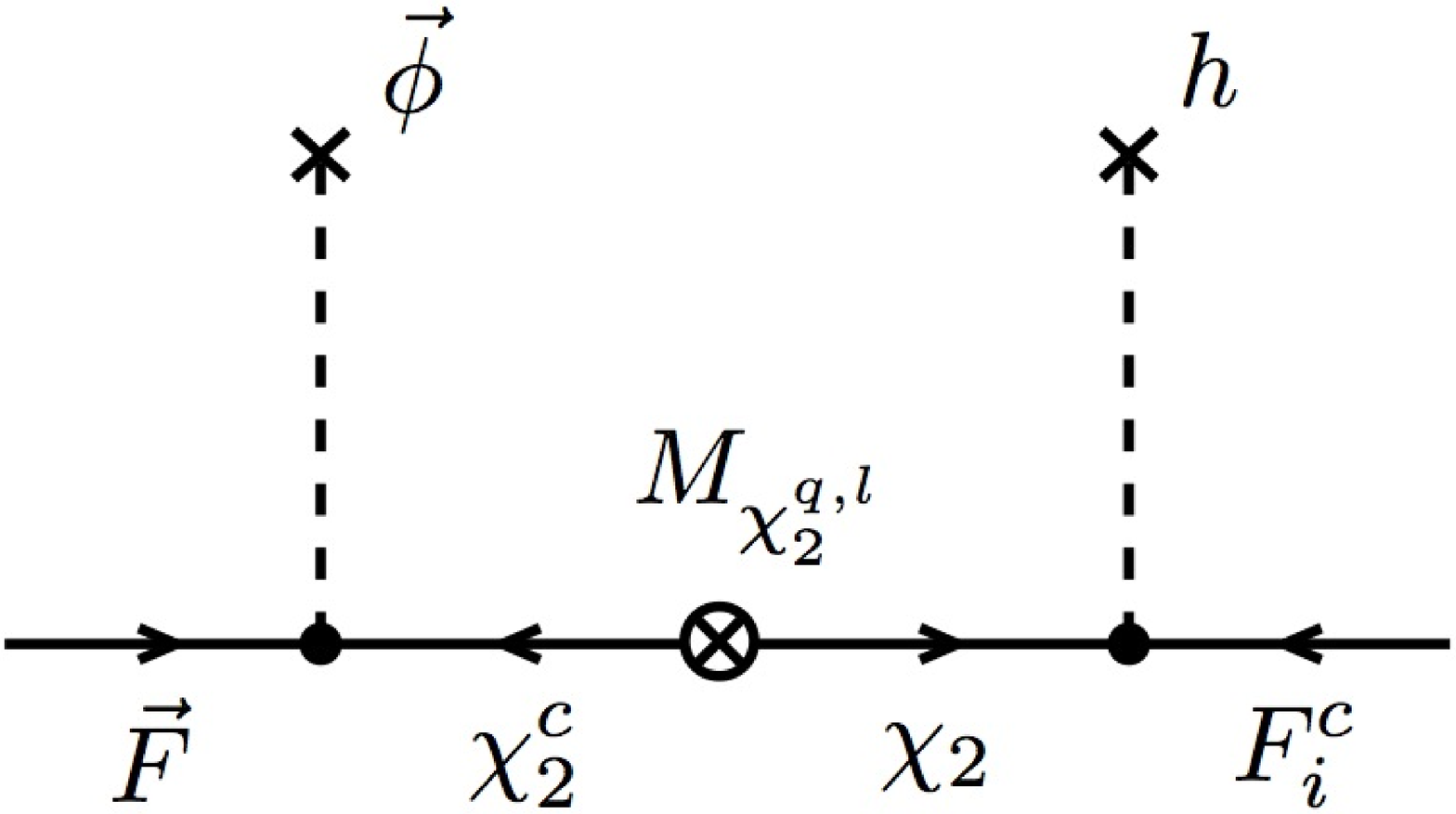}}
\ee
The main difference between them stems from the position of
the flavon and MSSM-like Higgs doublet insertions determining the
transformation properties of the relevant messenger fields
$\chi_1$ and $\chi_2$. The Pati-Salam quantum numbers of
$\chi_{1}$ are $(4,1,2)$ while $\chi_{2}$ transforms as $(4,2,1)$.
The $\chi_{1}$ messenger is ``universal'' as it feels only the
quantum numbers of $\vec F$ and $h$ while $\chi_{2}$ is ``flavon
specific'', because it carries the flavon extra charges (and thus
should be called $\chi_{2}^{\phi}$).

Upon the Pati-Salam spontaneous symmetry breaking the $\chi_{1}$ multiplet splits into four
distinct states (denoted in what follows by superscripts
$u,d,e,\nu$) while $\chi_{2}^{\phi}$ decays into just 2 states
because it does not feel the $SU(2)_{L}$ charges of the quarks and
leptons (that is why it is furnished by only a pair of
superscripts $q,l$). Playing with masses of $\chi_{1}^{u,d,e,\nu}$
one affects uniformly all the linear terms within a specific
Yukawa matrix, while adjusting the masses of $\chi_{2}^{\phi;q,l}$
leads to changes of entries generated by the appropriate flavons
(the specific $\chi_{2}^\phi$ is associated to), but without
differences between those in the up and down and charged lepton
and neutrino sectors respectively.

Remarkably enough, the minimal set of messengers leading to
potentially realistic quark and lepton Dirac Yukawa textures is
very concise, as shown in Table \ref{messengers1}.
\begin{table}[ht]
\centering
\begin{tabular}{|c|c|c|c|c|}
\hline
field & $SU(4)\otimes SU(2)_{L}\otimes SU(2)_{R}$ & $SO(3)$ & $U(1)$  & $Z_{2}$\\
\hline
$\chi_{1}$, $\chi_{1}^{c}$&  $(4,1,2)$, $(\overline{4},1,2)$ & $3$ & 0 & $ + $\\
$\tilde{\chi}_{1}$, $\tilde{\chi}_{1}^{c}$&  $(4,1,2)$, $(\overline{4},1,2)$ & $1$ & 0 & $ - $\\
$\chi_{2}$, $\chi_{2}^{c}$ &  $(4,2,1)$, $(\overline{4},2,1)$  & $1$ & $\pm 3$ & $ - $\\
\hline
 $\psi,\overline{\psi}$ & $(\overline{10},1,3)$,\,$({10},1,3)$ & $1$ & $0$ & $+$ \\
 $\Psi$ & $(1,1,1)$ & $1$ & $0$ & $-$ \\
\hline
\end{tabular}
\caption{\label{messengers1} The ``level-1'' messenger sector of
the model  responsible for the desired Dirac Yukawa structures. }
\end{table}

The interactions of the relevant messengers with the
matter, flavon and Higgs fields are given by
$$ W_{\chi}= \vec{\chi}_{1}. (
y_{\chi_{1}F^{c}_{1}\phi_{23}}F_{1}^{c} \vec\phi_{23}+
y_{\chi_{1}F^{c}_{2}\phi_{123}}F_{2}^{c} \vec\phi_{123}+
y_{\chi_{1}F^{c}_{3}\phi_{3}}F_{3}^{c} \vec\phi_{3} )
+y_{\chi_{1}^{c}Fh}\vec{F}.\vec{\chi}_{1}^{c}h
+y_{\chi_{1}\chi_{1}^{c}\phi_{12}}(\vec{\chi}_{1}\times
\vec{{\chi}}_{1}^{c}).\vec{{\phi}}_{12}  
$$
\be
+y_{\chi_{2}F^{c}_{3}h}\chi_{2}F_{3}^{c}h
+y_{\chi_{2}^{c}F\phi_{3}}\vec{F}.\vec\phi_{3}{\chi}_{2}^{c}
+y_{\tilde{\chi}_{1}\chi_{1}^{c}\tilde{\phi}_{23}}{\tilde{\chi}}_{1}
(\vec{\chi}_{1}^{c}.\vec{\tilde{\phi}}_{23})
+y_{\tilde{\chi_{1}}F^{c}_{2}\Sigma}{\tilde{\chi}}_{1}
F_{2}^{c}\Sigma \ee

\begin{figure}[ht]
\centering \bea
\parbox{3.5cm}{\includegraphics[width=3.5cm]{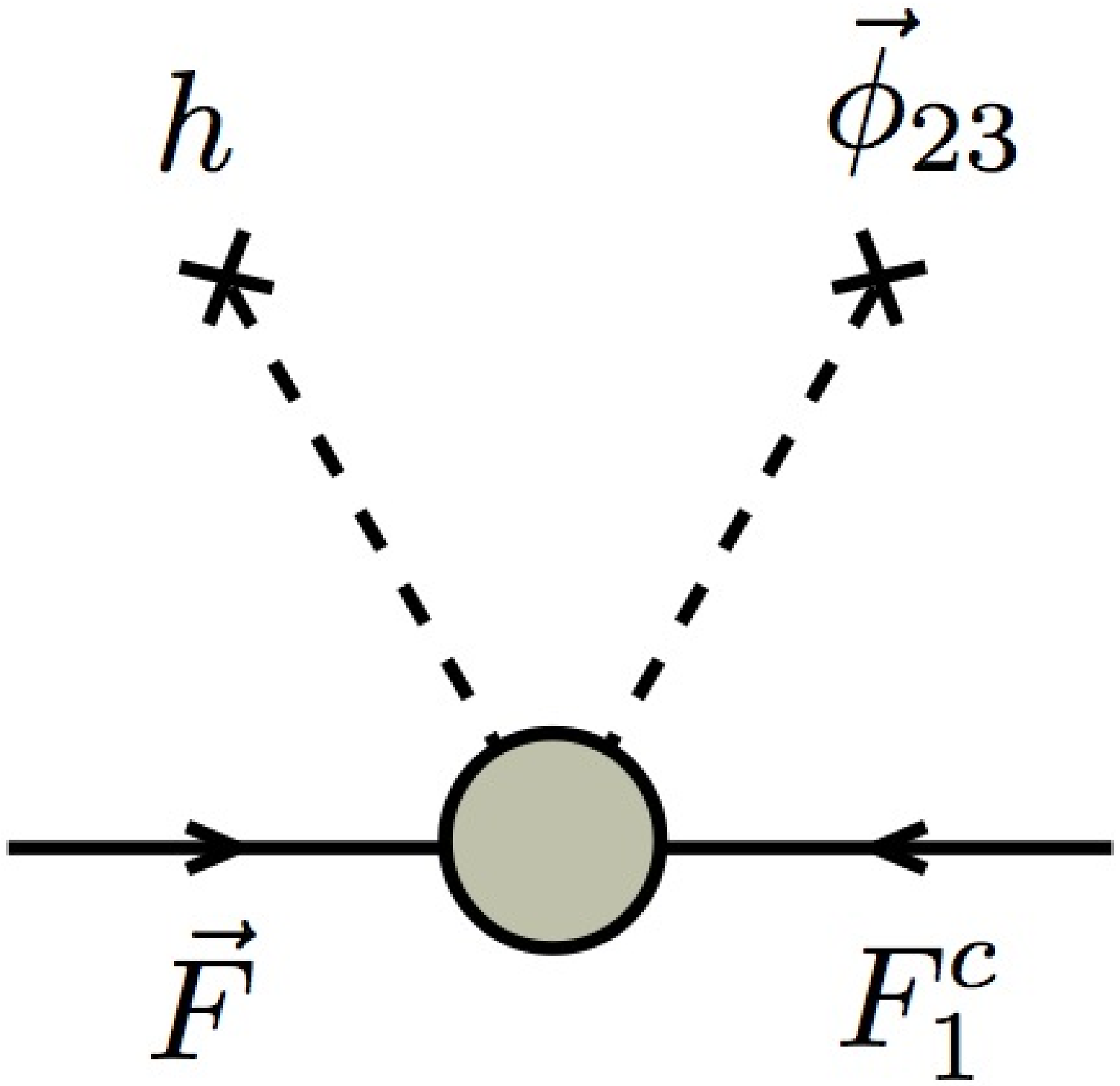}} & \lowpos{=} &
\parbox{5cm}{\includegraphics[width=5cm]{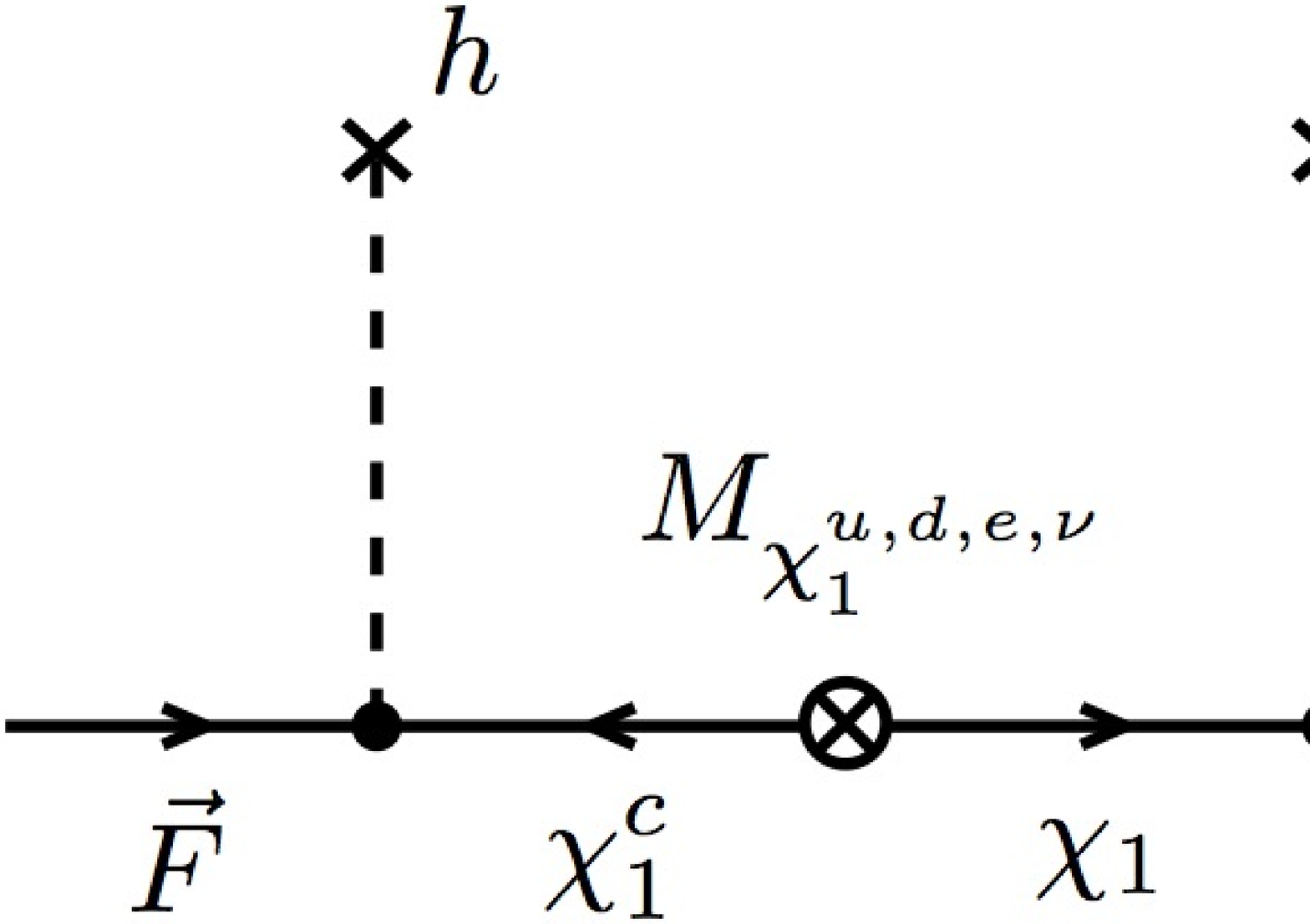}} \nonumber\\
\parbox{3.5cm}{\includegraphics[width=3.5cm]{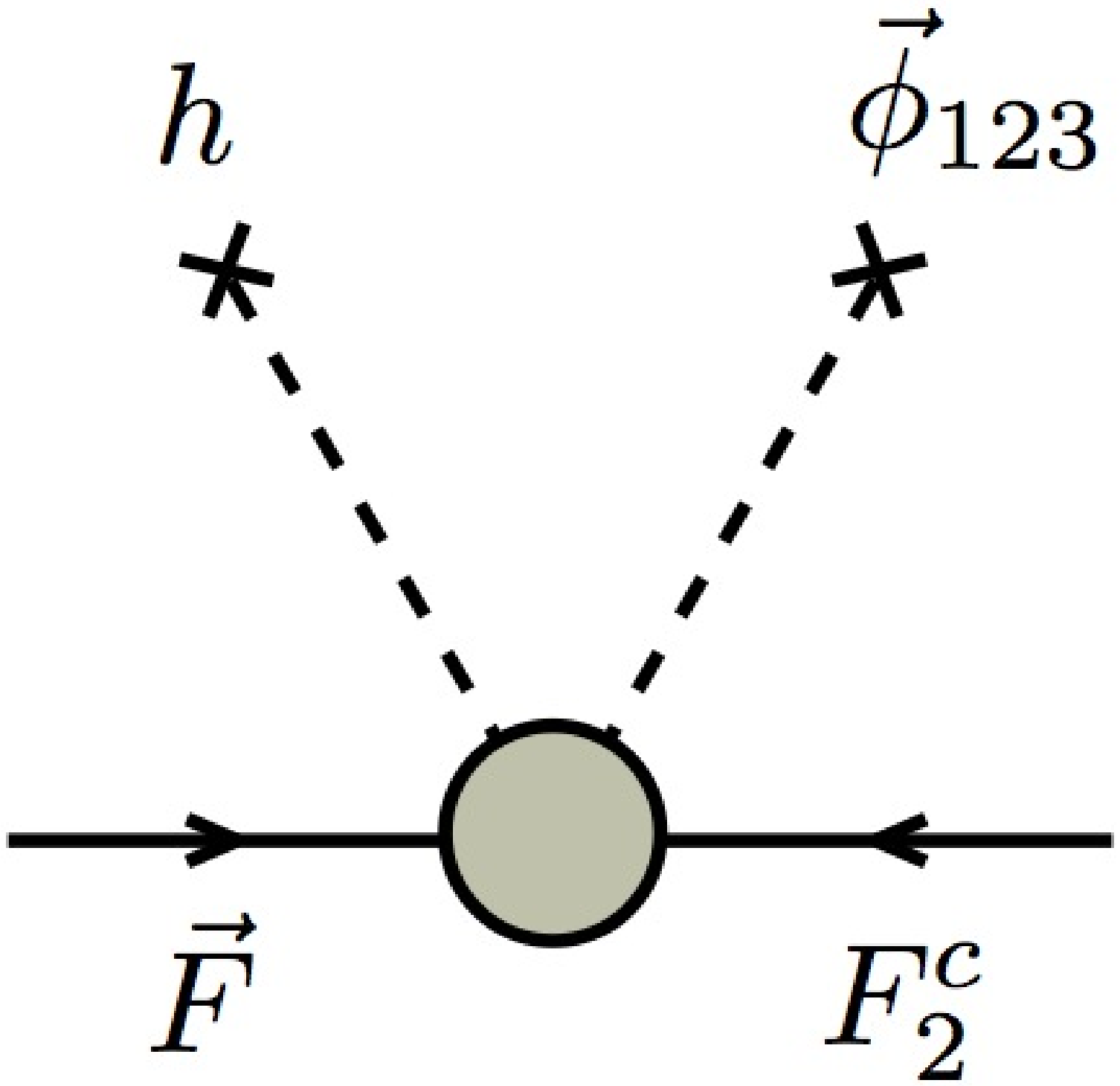}} & \lowpos{=} &
\parbox{5cm}{\includegraphics[width=5cm]{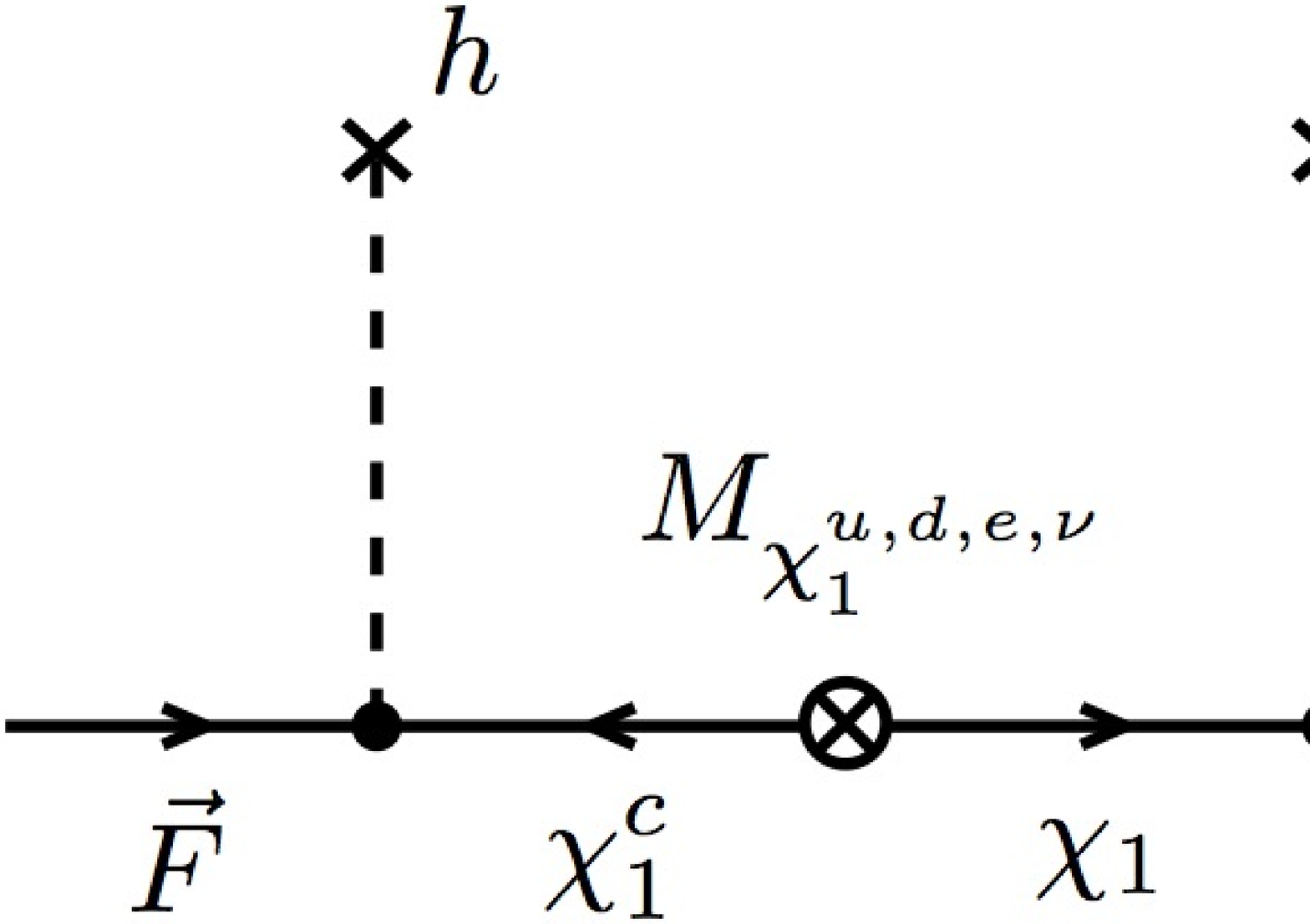}}\nonumber
\eea 
$$
\parbox{3.5cm}{\includegraphics[width=3.5cm]{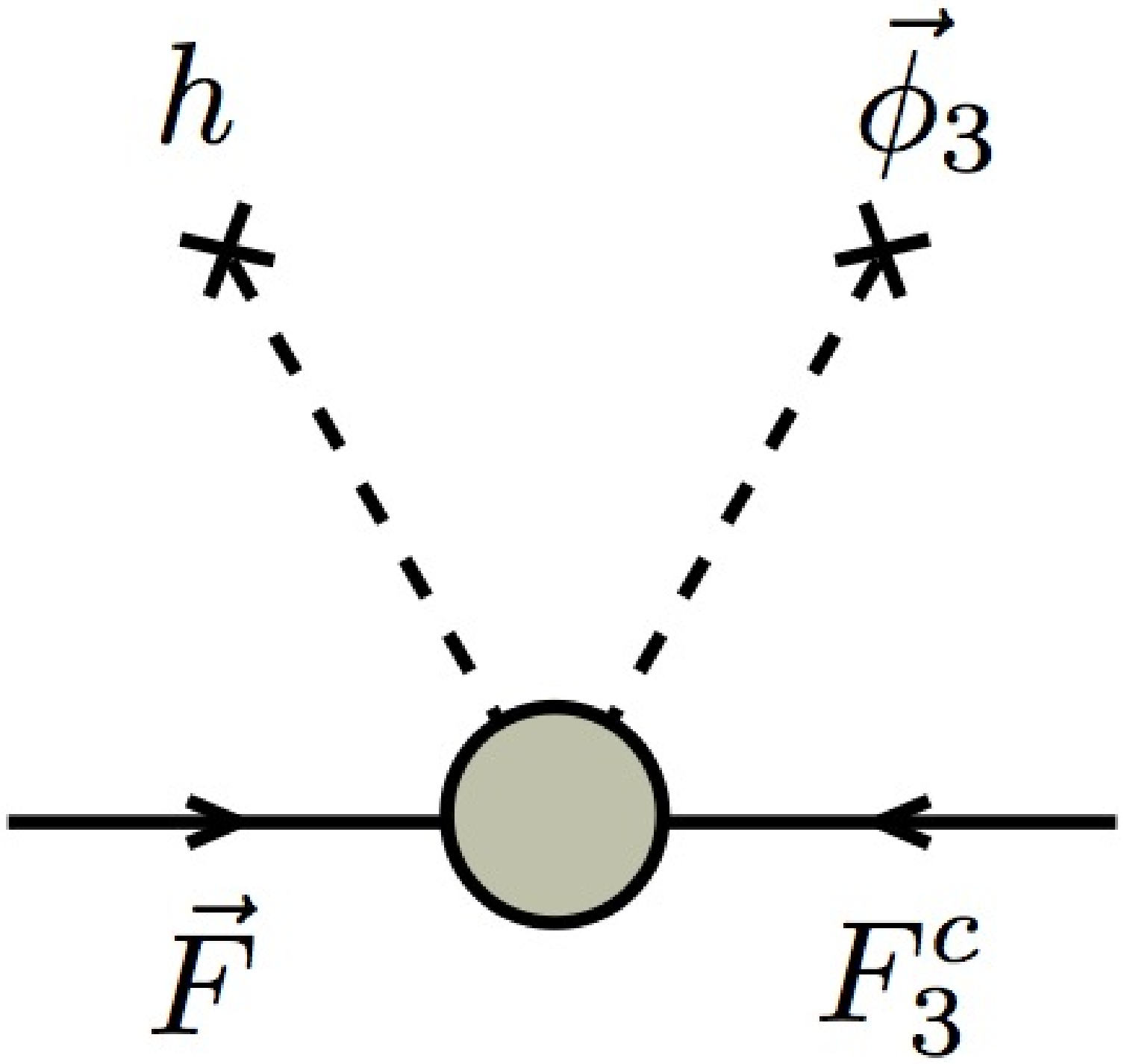}} \lowpos{=}
\parbox{5cm}{\includegraphics[width=5cm]{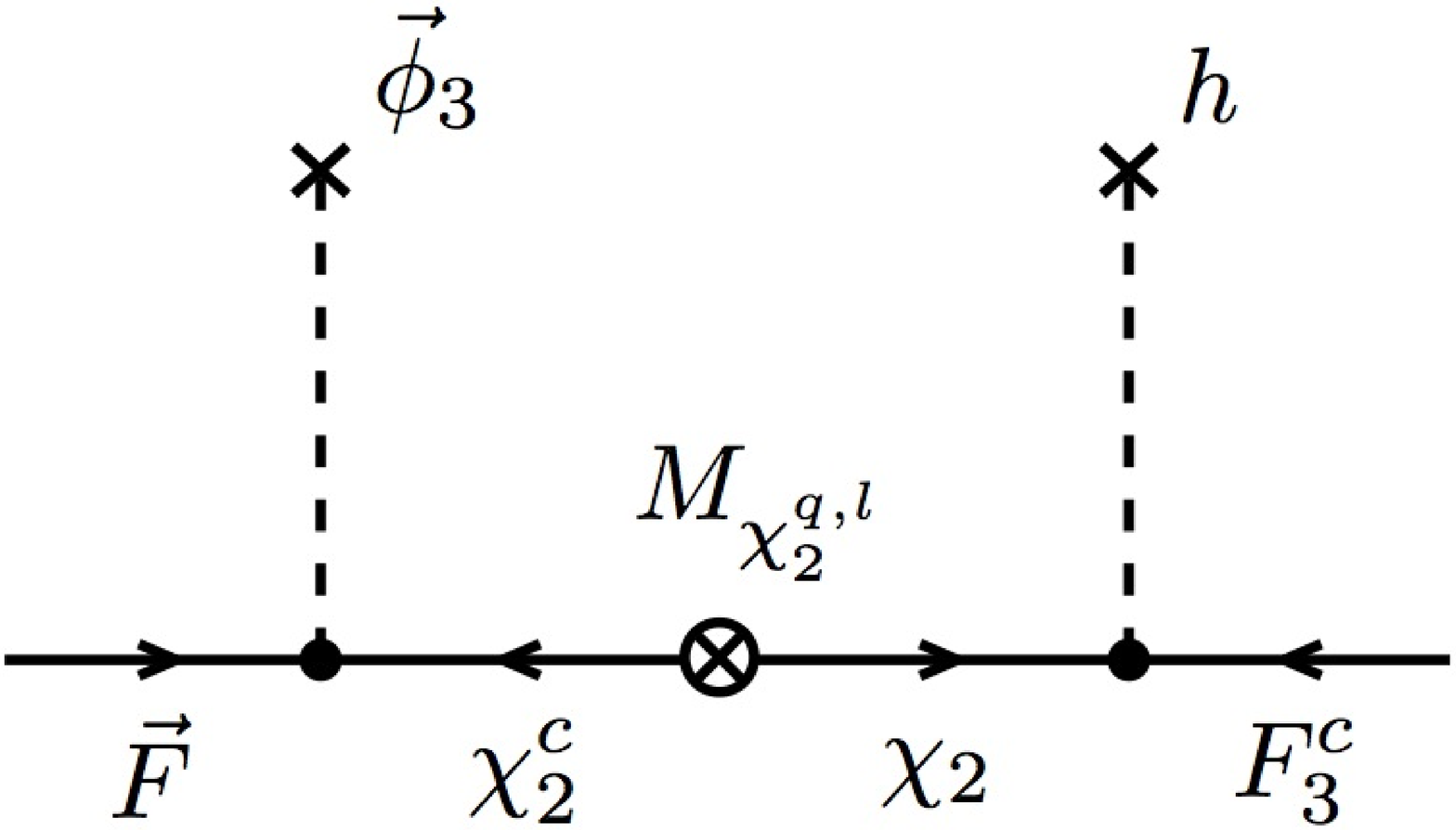}}\lowpos{+}
\parbox{5cm}{\includegraphics[width=5cm]{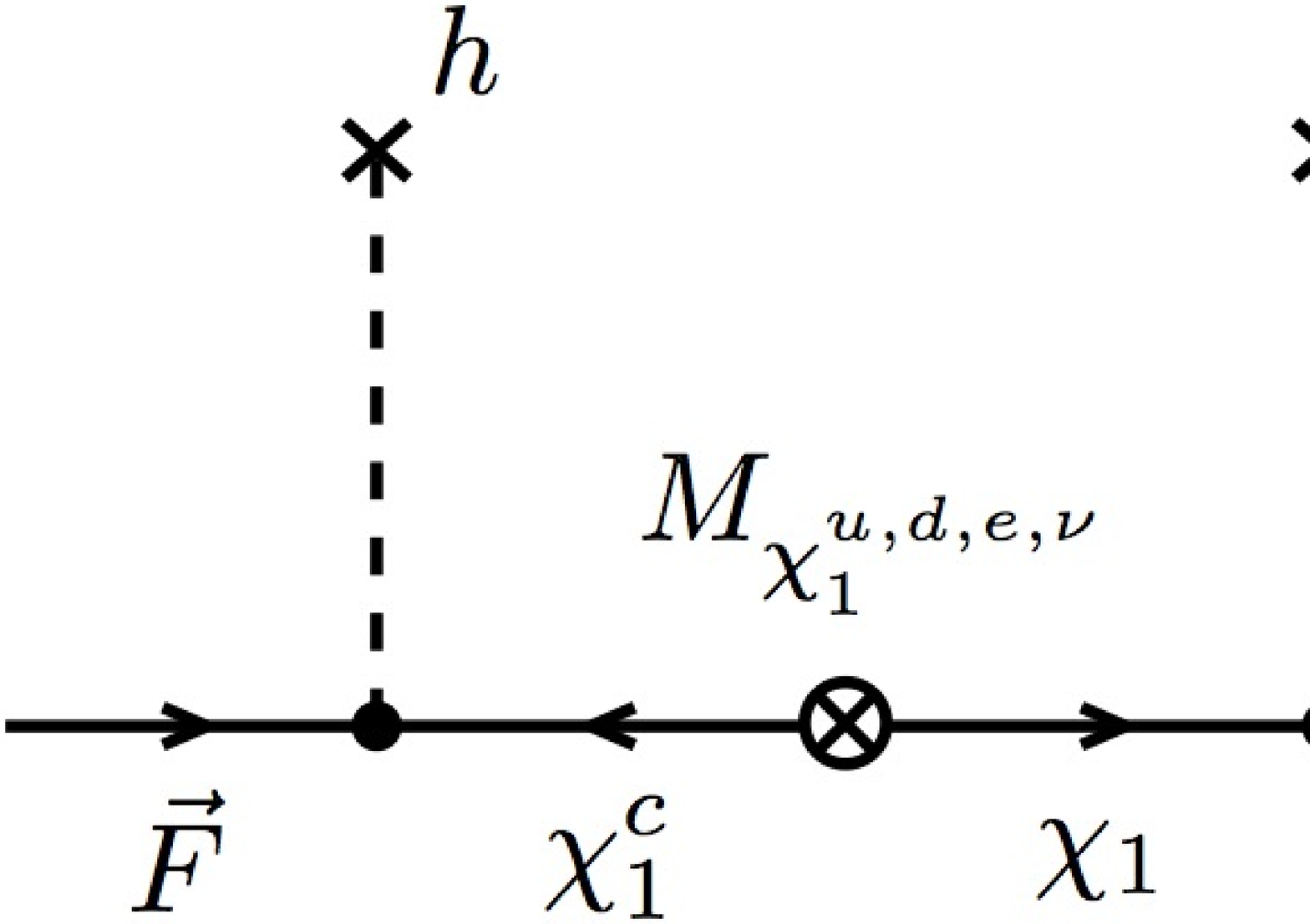}}
$$
\caption{The structure of a typical contributions to the Dirac
masses
  of matter fermions. Since the $\chi_{1}$ messenger does not ``feel''
  the extra quantum numbers of $F_{x}^{c}$ and $\vec{\phi}_{y}$ these
  topologies are generic for all the Dirac entries. On the other hand,
  the $U(1)\times Z_{2}$ charges of the $\chi_{2}$ messenger
  dominating the 33 entries are such that it can (at the
  lowest level) accompany  only the $\vec{\phi_{3}}$ flavon. Moreover, upon
  $SU(4)\otimes SU(2)_{L}\otimes SU(2)_{R}\to SU(3)_{c}\otimes
  SU(2)_{L}\otimes U(1)_{Y}$ it can split only into a pair of states
  $\chi_{2}^{q,l}$ giving rise to universal entries in the quark and
  lepton sectors respectively. However, in our setup this
  splitting can not emerge at the lowest level and thus the $b-\tau$
  Yukawa unification is preserved up to higher order corrections.}
\label{dirac1-basic}
\end{figure}

The internal structure of the lowest level Dirac operators is depicted in Fig.
\ref{dirac1-basic}. It is assumed that there is only one light
enough flavon specific messenger of type 2 (c.f. discussion below formula
(\ref{messengertypes})) associated with
$\vec\phi_{3}$.

\begin{figure}[ht]
\centering 
$$
\parbox{3.5cm}{\includegraphics[width=3.5cm]{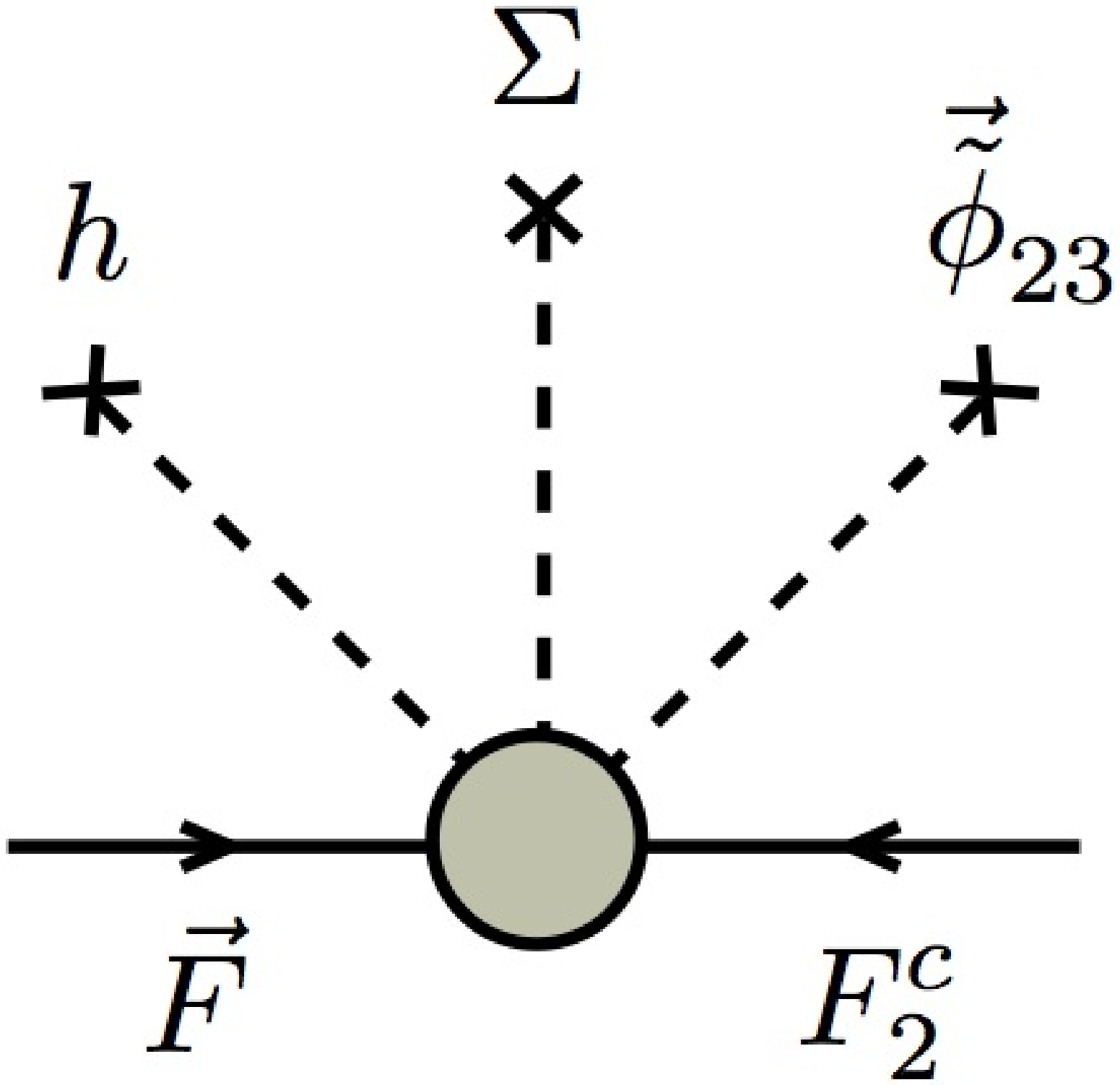}}\lowpos{=} \parbox{8cm}{\includegraphics[width=8cm]{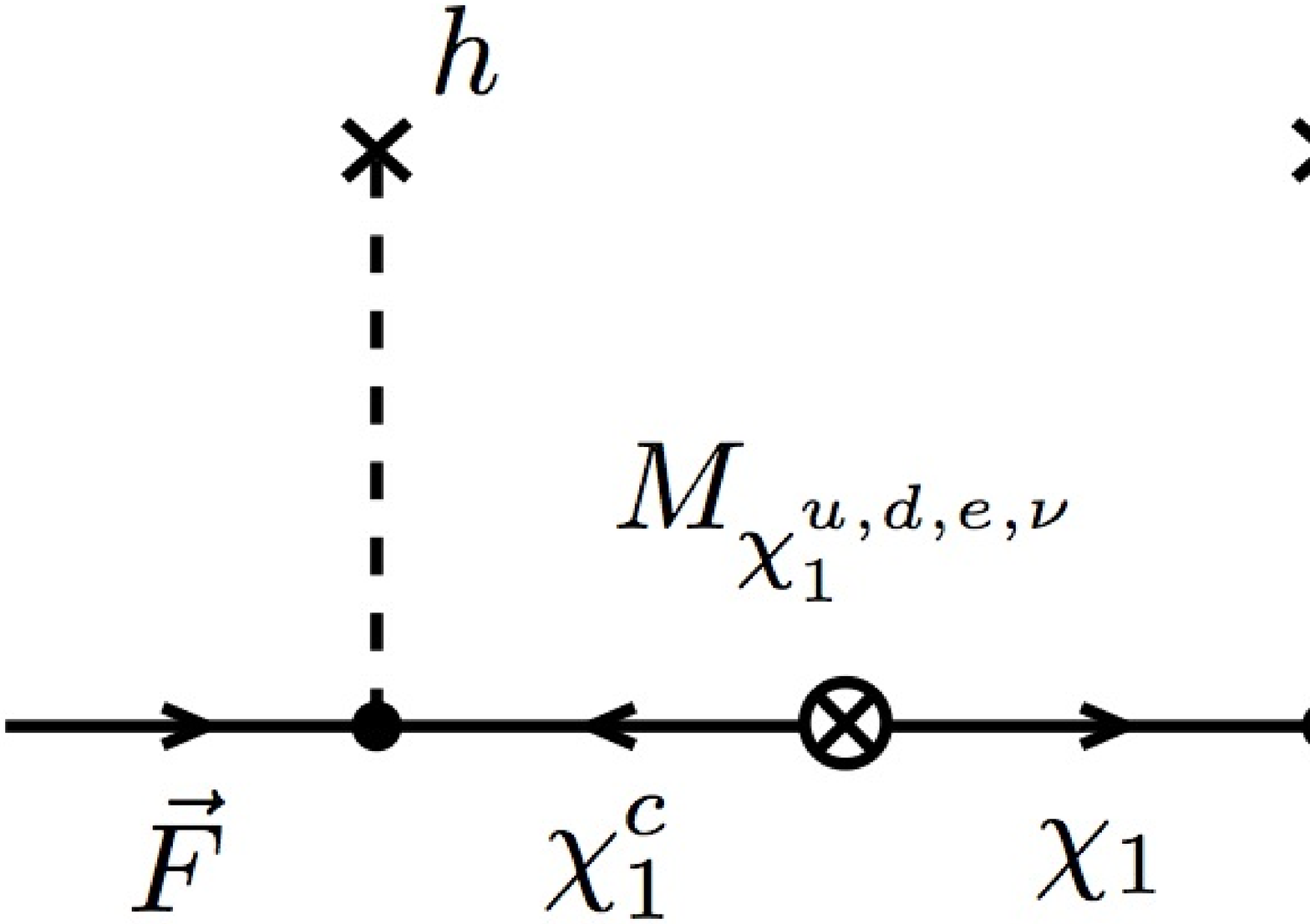}}
$$
  \caption{The structure of the Georgi-Jarlskog operator leading to the desired Clebsch-Gordon coefficient in the charged lepton sector giving the proper $m_{s}/m_{\mu}$ ratio. Since the projection of $\langle\Sigma\rangle$ in the $Y=0$ direction is zero, the tri-bimaximal mixing in the neutrino sector remains unaffected.}
\label{GJfigure}
\end{figure}

\begin{figure}[ht]
\centering $$
\parbox{3.5cm}{\includegraphics[width=3.5cm]{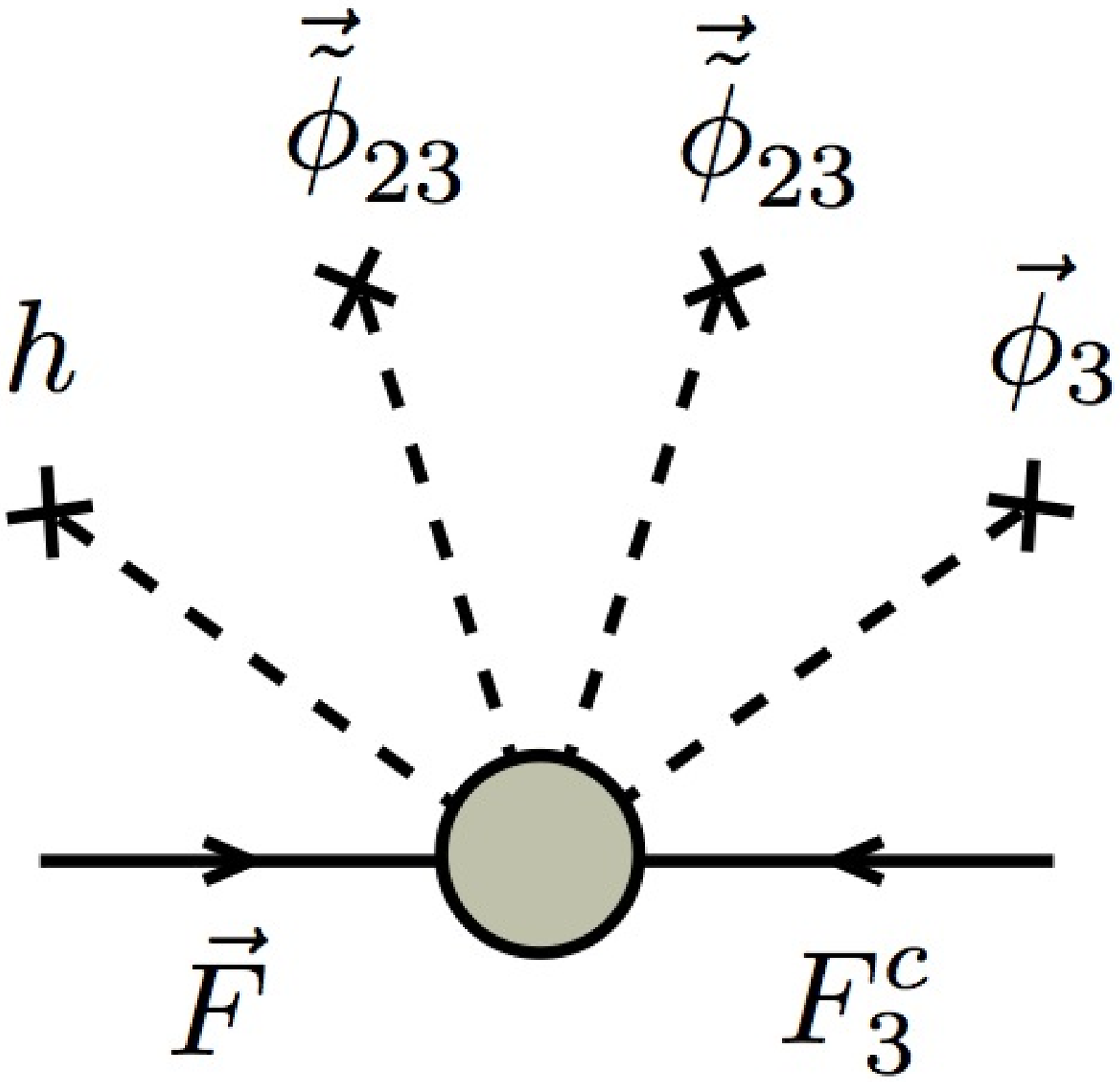}}\lowpos{=}
\parbox{11cm}{\includegraphics[width=11cm]{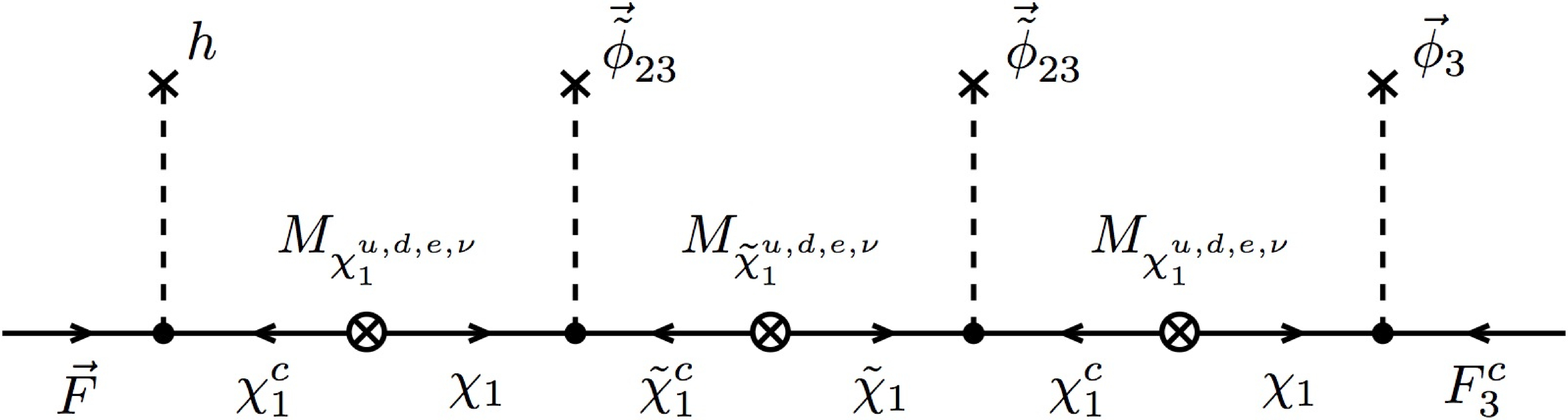}}
$$ $$
\parbox{3.5cm}{\includegraphics[width=3.5cm]{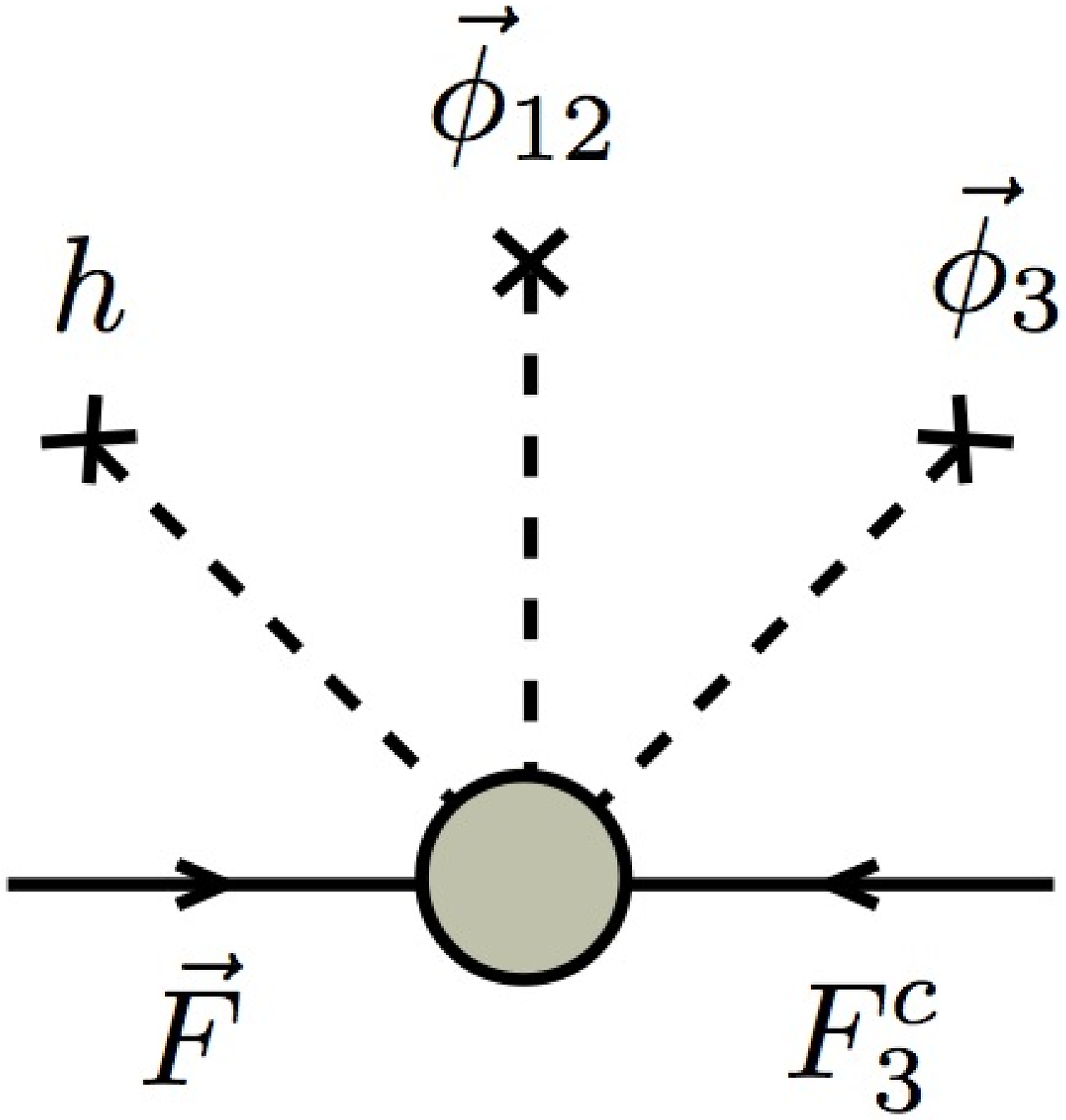}}\lowpos{=}
\parbox{8cm}{\includegraphics[width=8cm]{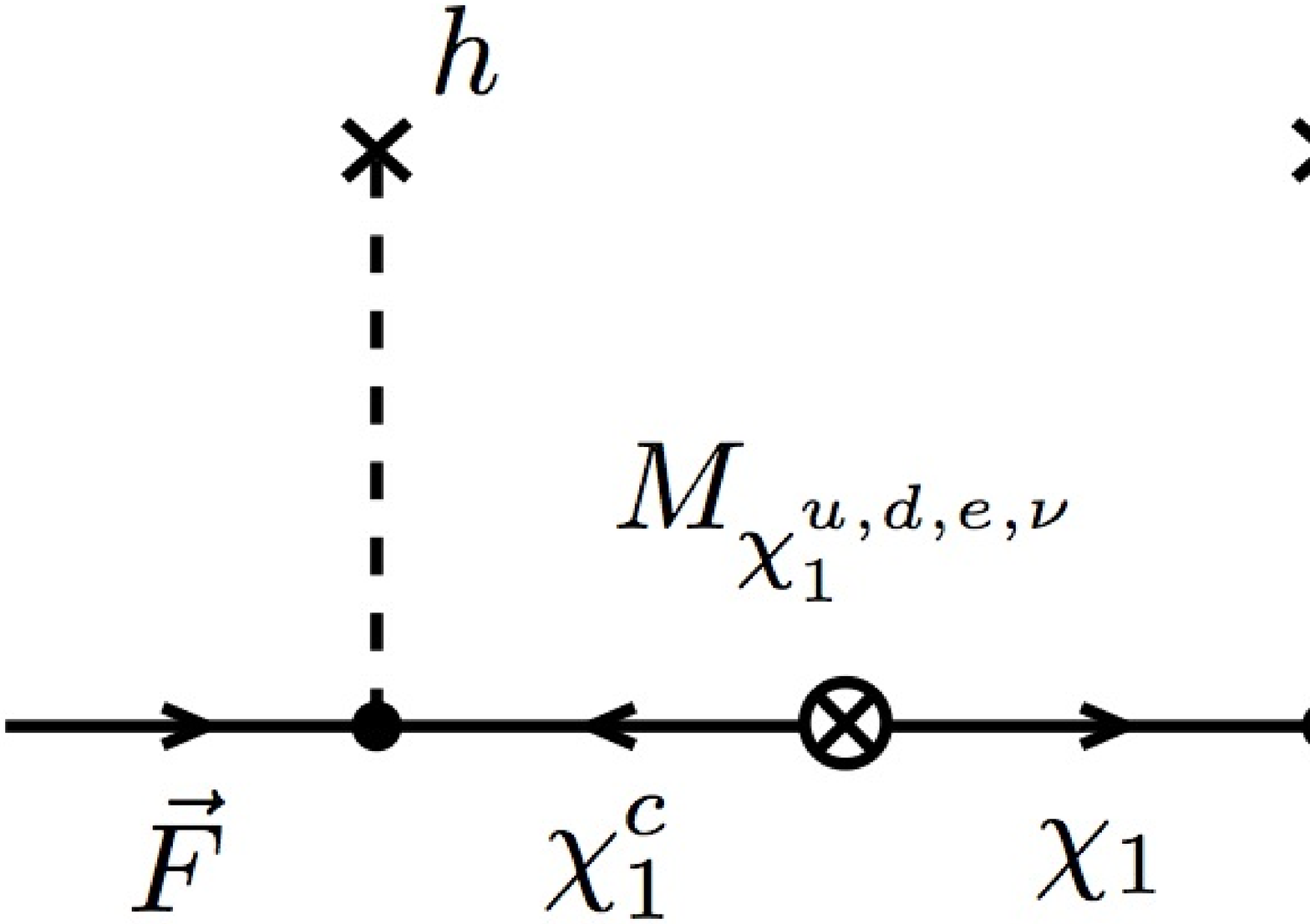}}
$$
  \caption{The leading contributions to the Yukawa 13 and 23 entries
  emerge automatically due to the particular $\chi_{1}$ and
  $\vec{\tilde{\phi}}_{23}$ quantum numbers. The $SO(3)$ indices are
  contracted so that the leftmost $\vec{\tilde{\phi}}_{23}$ couples to
  $\vec{F}$ and the $\vec{\tilde{\phi}}_{23}$ flavon on the right
  saturates the $SO(3)$ index of $\vec{\phi}_{3}$.  This is the only
  option since $\tilde{\chi}_{1}$ is an $SO(3)$ singlet.
\label{23figure}}
\end{figure}

%----
\newpage
The structure of the higher order operators responsible for the
Georgi-Jarlskog structure (requiring an extra type-1 messenger
field denoted by $\tilde\chi_1$) and the 1-3 and 2-3 CKM mixings
is shown in Figs. \ref{GJfigure} and \ref{23figure}.
%----
%------------------------------------------
\subsection{The Yukawa matrices\label{Yukawa}}
%------------------------------------------
With the messenger sector specified, we can now return to the
Dirac Yukawa matrix structure in Eq.\ref{diracpart}, and express the
separate Yukawa matrices in each charge sector in terms of
the messenger masses. The relevant expressions read (dropping the $LR$ subscript for simplicity):
\be
\label{Yu} Y^{u}= \left(\begin{array}{ccc}
0 &y_{123} \varepsilon^{u}_{123}&  {y}_{12} \varepsilon^{u}_{12} \varepsilon^{u}_{3}\\
y_{23} \varepsilon^{u}_{23} & y_{123} \varepsilon^{u}_{123}  -2  y_{GJ}\tilde{\varepsilon}^{u}_{23}\sigma^u & \tilde{y}_{23}(\tilde{\varepsilon}^{u}_{23})^{2} \varepsilon^{u}_{3}\\
-y_{23}\varepsilon^{u}_{23} & y_{123} \varepsilon^{u}_{123}  -2
y_{GJ}\tilde{\varepsilon}^{u}_{23}\sigma^u
&y_{3}'\varepsilon^{q}_{3}+ y_{3}\varepsilon^{u}_{3}
\end{array}\right)
\ee
\be
\label{Yd}
Y^{d}= \left(\begin{array}{ccc}
0 &y_{123} \varepsilon^{d}_{123} &  {y}_{12} \varepsilon^{d}_{12} \varepsilon^{d}_{3} \\
y_{23} \varepsilon^{d}_{23} & y_{123} \varepsilon^{d}_{123}+  y_{GJ}\tilde{\varepsilon}^{d}_{23}\sigma^d & \tilde{y}_{23}(\tilde{\varepsilon}^{d}_{23})^{2} \varepsilon^{d}_{3} \\
-y_{23}\varepsilon^{d}_{23} &y_{123} \varepsilon^{d}_{123} +
y_{GJ}\tilde{\varepsilon}^{d}_{23}\sigma^d &
y_{3}'\varepsilon^{q}_{3}+y_{3}\varepsilon^{d}_{3}
\end{array}\right)
\ee
\be
\label{Ye}
Y^{e}= \left(\begin{array}{ccc}
0 &y_{123} \varepsilon^{e}_{123}&  {y}_{12} \varepsilon^{e}_{12} \varepsilon^{e}_{3}\\
y_{23} \varepsilon^{e}_{23} & y_{123} \varepsilon^{e}_{123}   +3  y_{GJ}\tilde{\varepsilon}^{e}_{23}\sigma^e & \tilde{y}_{23}(\tilde{\varepsilon}^{e}_{23})^{2} \varepsilon^{e}_{3}\\
-y_{23}\varepsilon^{e}_{23} & y_{123} \varepsilon^{e}_{123} +3
y_{GJ}\tilde{\varepsilon}^{e}_{23}\sigma^e
&y_{3}'\varepsilon^{l}_{3}+ y_{3}\varepsilon^{e}_{3}
\end{array}\right)
\ee
\be
\label{Ynu}
Y^{\nu}= \left(\begin{array}{ccc}
0 &y_{123} \varepsilon^{\nu}_{123} &  {y}_{12} \varepsilon^{\nu}_{12} \varepsilon^{\nu}_{3}\\
y_{23} \varepsilon^{\nu}_{23}  & y_{123} \varepsilon^{\nu}_{123} & \tilde{y}_{23}(\tilde{\varepsilon}^{\nu}_{23})^{2} \varepsilon^{\nu}_{3}\\
-y_{23}\varepsilon^{\nu}_{23} & y_{123} \varepsilon^{\nu}_{123} &
y_{3}'\varepsilon^{l}_{3}+y_{3}\varepsilon^{\nu}_{3}
\end{array}\right)
\ee
where we have used the following abbreviations ($f$ stands for $u$,
$d$, $\nu$ and $e$):
\be
\varepsilon_{23}^{f}\equiv
\frac{|\vev{\vec{\phi}_{23}}|}{M_{\chi_{1}^{f}}}, \quad
\varepsilon_{123}^{f}\equiv
\frac{|\vev{\vec{\phi}_{123}}|}{M_{\chi_{1}^{f}}},\quad
\varepsilon_{3}^{f}\equiv
\frac{|\vev{\vec{\phi}_{3}}|}{M_{\chi_{1}^{f}}},\quad
\varepsilon_{12}^{f}\equiv
\frac{|\vev{\vec{\phi}_{12}}|}{M_{\chi_{1}^{f}}},\quad
\tilde{\varepsilon}_{23}^{f}\equiv
\frac{|\vev{\vec{\tilde{\phi}}_{23}}|}{M_{\chi_{1}^{f}}}
\ee
\be
\varepsilon_{3}^{q}\equiv
\frac{|\vev{\vec{\phi}_{3}}|}{M_{\chi_{2}^{q}}},\quad
\varepsilon_{3}^{l}\equiv
\frac{|\vev{\vec{\phi}_{3}}|}{M_{\chi_{2}^{l}}}\quad {\rm and} \quad
\sigma^f\equiv \frac{|\vev{\Sigma}|}{M_{\chi_{1}^{f}}}
\ee

The above effective Yukawa matrices are obtained upon
integrating out the heavy messenger sector,
leading to the following relations between the dimensionless couplings:
\be\label{couplings} y_{23}\equiv
y_{\chi_{1}^{c}Fh}\, y_{\chi_{1}F^{c}_{1}\phi_{23}},\quad
y_{123}\equiv y_{\chi_{1}^{c}Fh}\,
y_{\chi_{1}F^{c}_{2}\phi_{123}},\quad y_{3}\equiv
y_{\chi_{1}^{c}Fh}\, y_{\chi_{1}F^{c}_{3}\phi_{3}},\quad
y_{3}'\equiv y_{\chi_{2}F^{c}_{3}h}\, y_{\chi_{2}^{c}F\phi_{3}},
\ee
$$
y_{GJ}\equiv y_{\chi_{1}^{c}Fh}\,
y_{\tilde{\chi}_{1}\chi_{1}^{c}\tilde{\phi}_{23}}\,
y_{\tilde{\chi}_{1}F^{c}_{2}\Sigma},\quad \tilde{y}_{23}\equiv
y_{\chi_{1}^{c}Fh}\,
y^{2}_{\tilde{\chi}_{1}\chi_{1}^{c}\tilde{\phi}_{23}}\,
y_{\chi_{1}F^{c}_{3}\phi_{3}},\quad {y}_{12}\equiv
y_{\chi_{1}^{c}Fh}\, y_{{\chi}_{1}\chi_{1}^{c}{\phi}_{12}}\,
y_{\chi_{1}F^{c}_{3}\phi_{3}}
$$

%------------------------------------------
\subsection{The messenger masses\label{messenger-spectra}}
%------------------------------------------

It is known that
the quark masses and mixing angles are well described by the following
textures \cite{Roberts:2001zy}:
\be
\label{Roberts}
|Y^{u}|\sim \left(\begin{array}{ccc}
0 & \varepsilon^{3} & O(\varepsilon^{3}) \\
\varepsilon^{3} & \varepsilon^{2} & O(\varepsilon^{2}) \\
O(\varepsilon^{3}) & O(\varepsilon^{2}) &1
\end{array}\right)
,\qquad |Y^{d}|\sim \left(\begin{array}{ccc}
0 & 1.5\overline{\varepsilon}^{3} & 0.4\overline{\varepsilon}^{3} \\
1.5\overline{\varepsilon}^{3} & \overline{\varepsilon}^{2} &
1.3\overline{\varepsilon}^{2} \\
O(\overline{\varepsilon}^{3}) & O(\overline{\varepsilon}^{2}) &1
\end{array}\right)
%,\qquad |Y^{l}|\sim \left(\begin{array}{ccc}
%0 & \overline{\varepsilon}^{3} &\overline{\varepsilon}^{3} \\
%\overline{\varepsilon}^{3} & 3 \overline{\varepsilon}^{2} & \overline{\varepsilon}^{2} \\
%\overline{\varepsilon}^{3} & \overline{\varepsilon}^{2} &1
%\end{array}\right)
,\qquad \ee with $\varepsilon\sim 0.05$ and
$\overline{\varepsilon}\sim 0.15$.
The charged lepton Yukawa matrix receives a form
similar to the down quark Yukawa matrix, but with a
``Georgi-Jarlskog'' factor of 3 in the $(2,2)$ entry of the
charged lepton matrix.

There is clearly a need to generate a
sizeable splitting in the spectrum of the $\chi_{1}$-type
messengers, in particular among the components coupled to the up
and down matter sectors. If we intend to reproduce the hierarchies
suggested by textures (\ref{Roberts}) the ratio of the down and
up $\chi_1$-type messenger masses
${M_{\chi_{1}^{d}}}/{M_{\chi_{1}^{u}}}\equiv r$ should be roughly
$\bar\varepsilon^3/{\varepsilon}^3\sim 1/30$. Thus, we should make
$M_{\chi_{1}^{u}}$ much heavier than $M_{\chi_{1}^{d}}$. Clearly,
this is possible only if the common bare masses in the
superpotential do not dominate the $\chi_{1}$ mass formula,
otherwise we get always $r\to 1$. It is also insufficient to split
the $\chi_{1}$ by means of Clebsch-Gordon coefficients of an extra
$\Sigma$-like Higgs field (in analogy with the Georgi-Jarlskog
mechanism) because $|r|$ in such a case is confined between the
minimum and maximum ratio of the relevant Clebsh-Gordon
coefficients (${\cal O}(1)$ numbers). Thus, we need an
alternative mechanism giving mass to $\chi_{1}^{u}$ only without
touching $\chi_{1}^{d}$.

\begin{figure}[ht]
\centering \bea
\parbox{3.5cm}{\includegraphics[width=3.5cm]{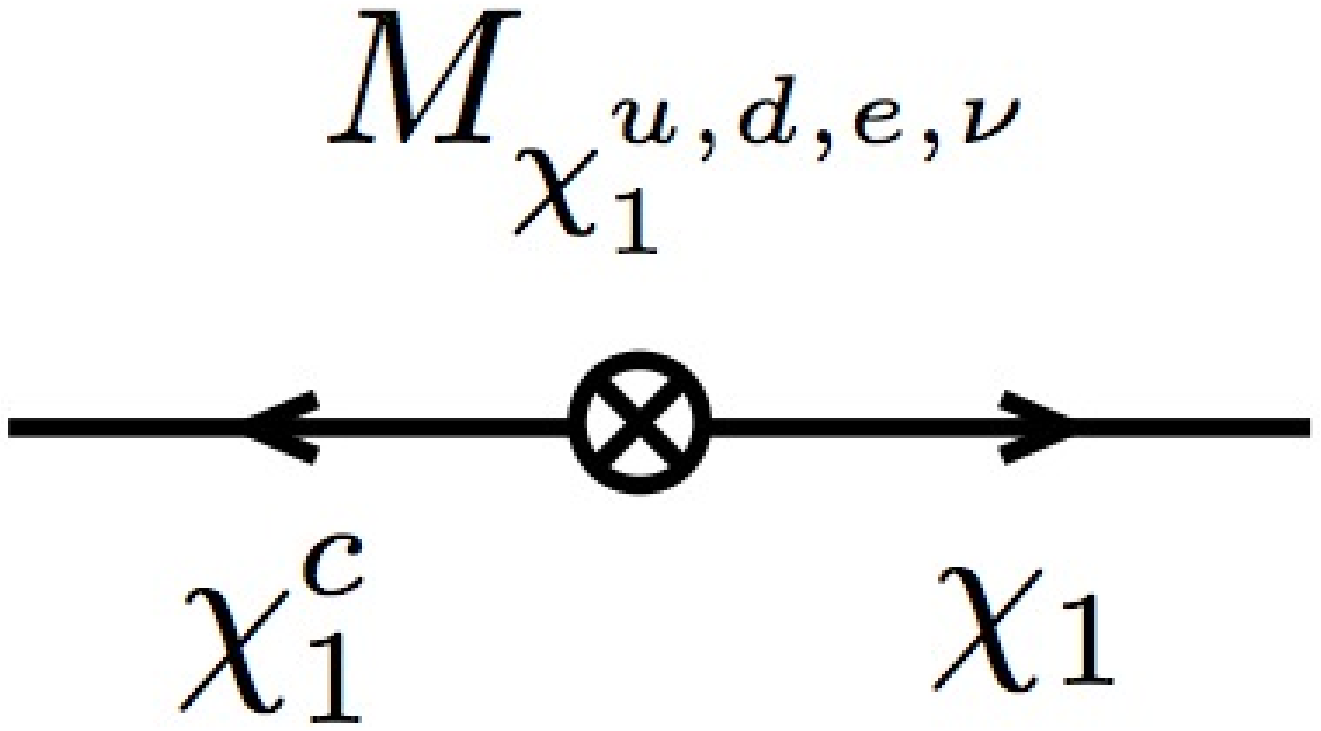}}
& \lowpos{=} &
\parbox{3.5cm}{\includegraphics[width=3.5cm]{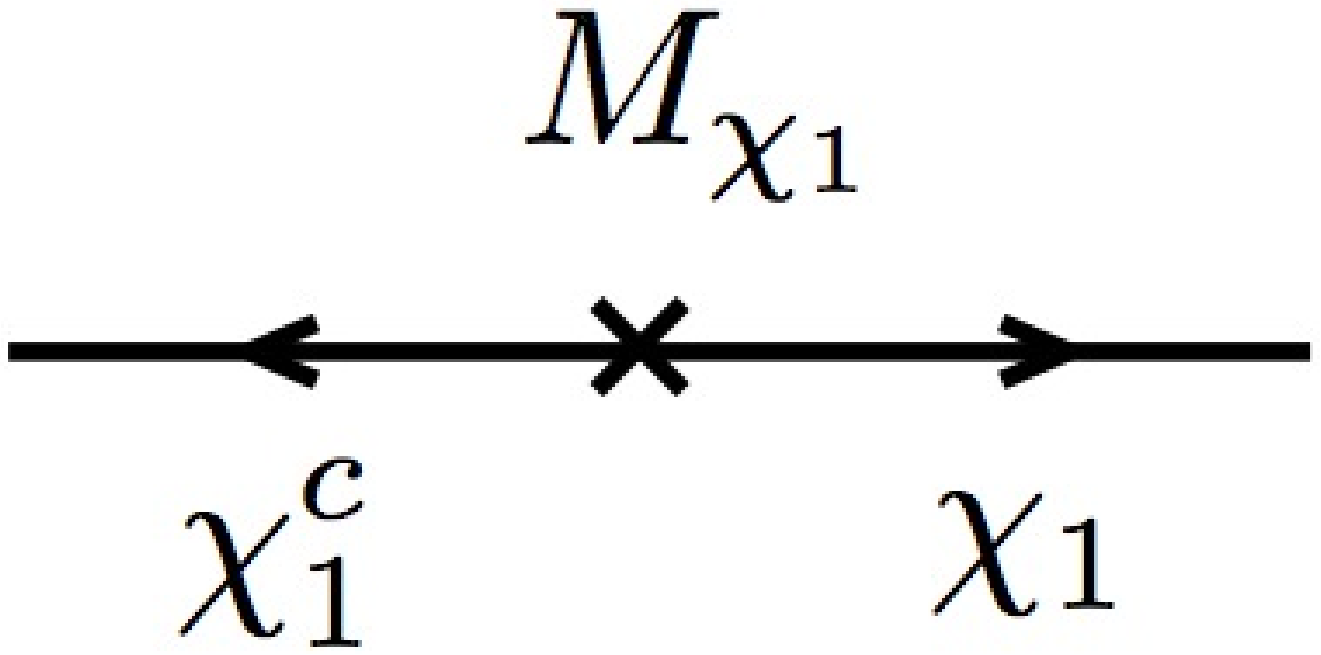}}
\lowpos{+}
\parbox{5cm}{\includegraphics[width=5cm]{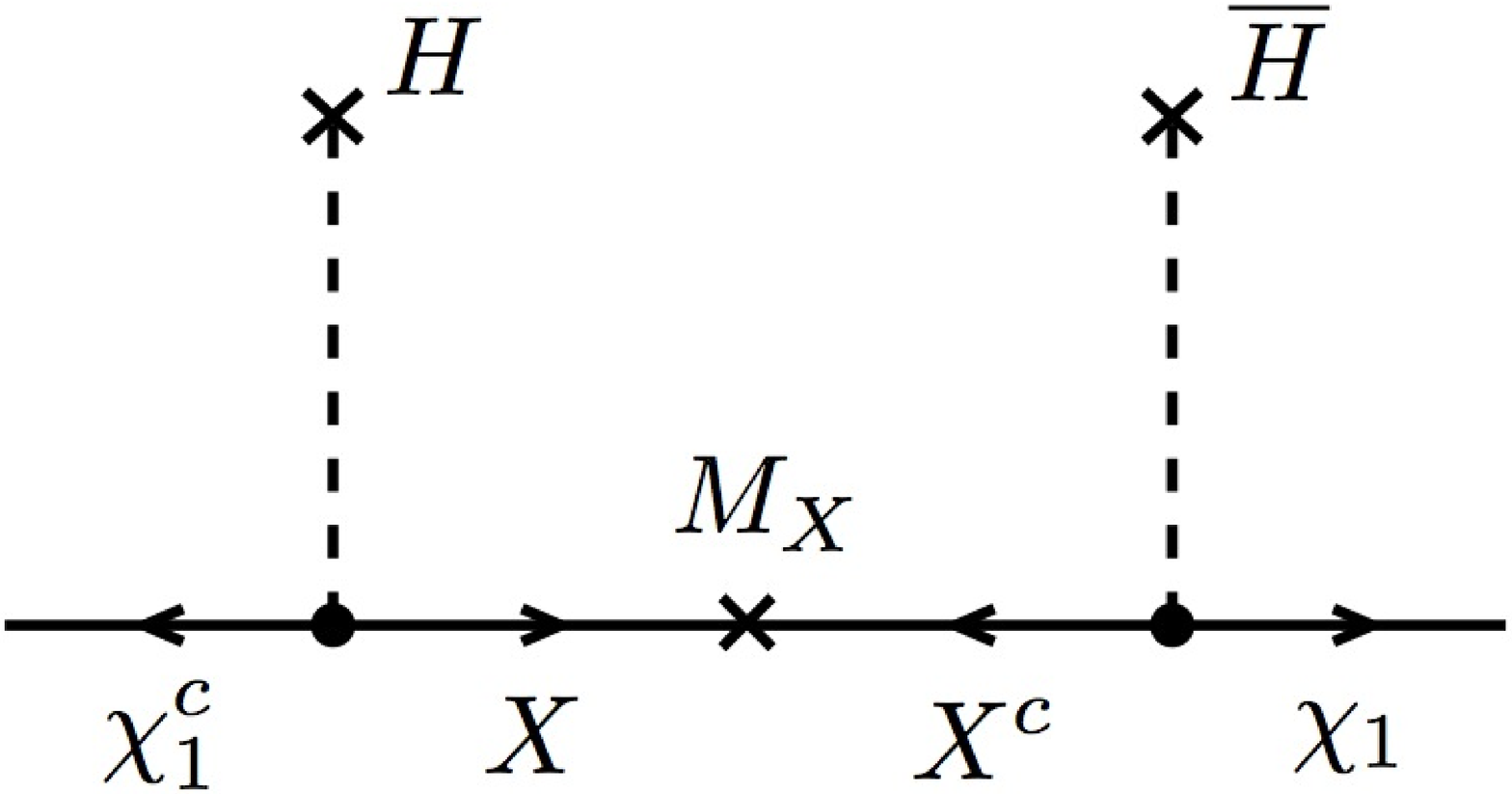}}
\lowpos{+ \ldots} \nonumber\\
\parbox{3.5cm}{\includegraphics[width=3.5cm]{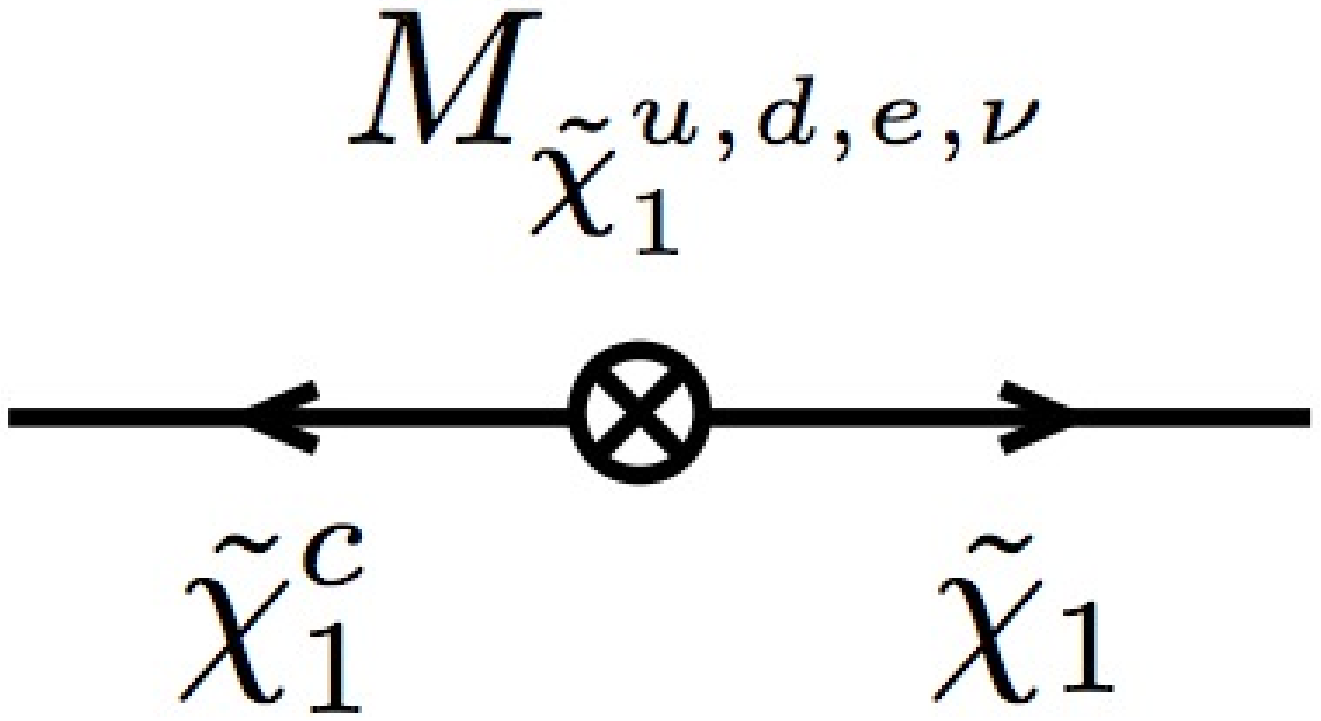}}
& \lowpos{=} &
\parbox{3.5cm}{\includegraphics[width=3.5cm]{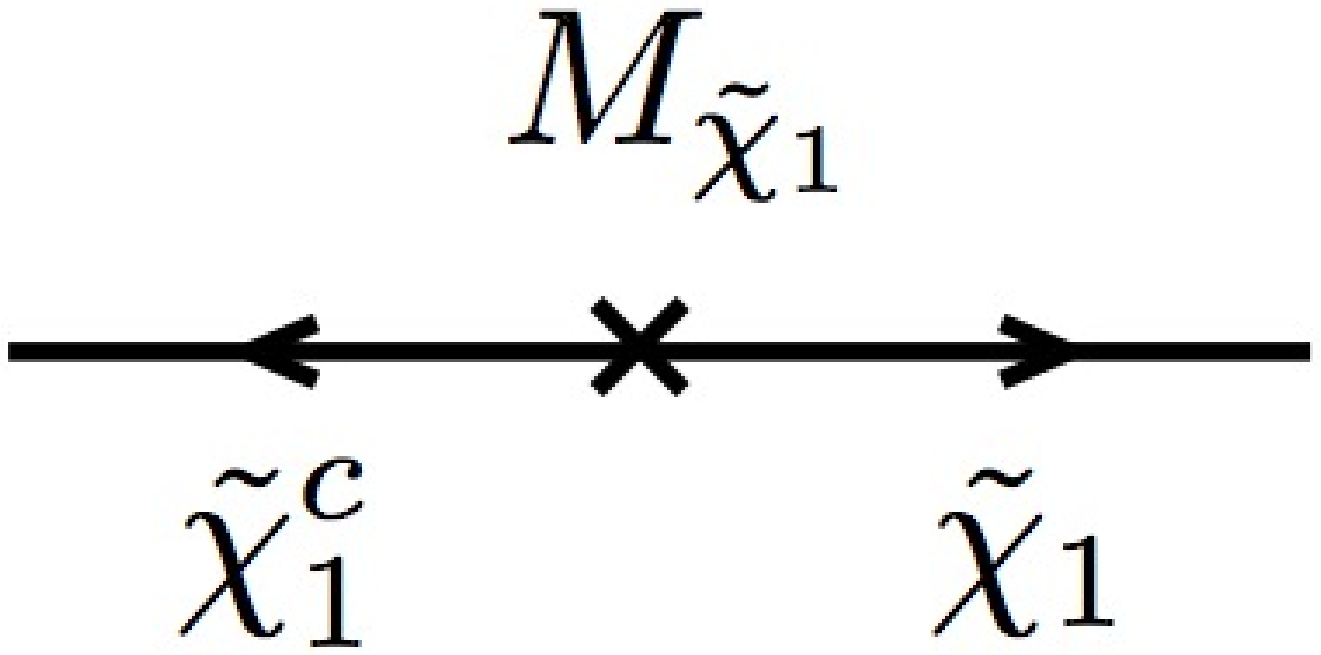}}
\lowpos{+ \ldots} \nonumber\\
\parbox{3.5cm}{\includegraphics[width=3.5cm]{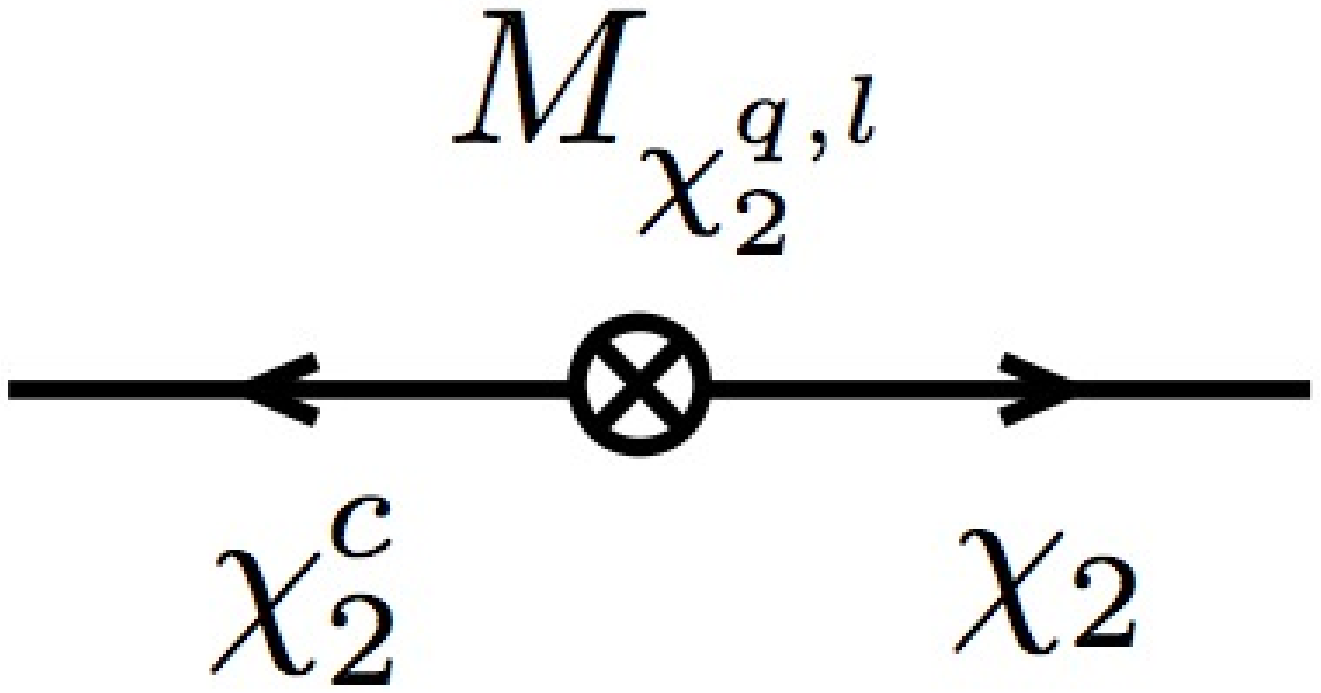}}
& \lowpos{=} &
\parbox{3.5cm}{\includegraphics[width=3.5cm]{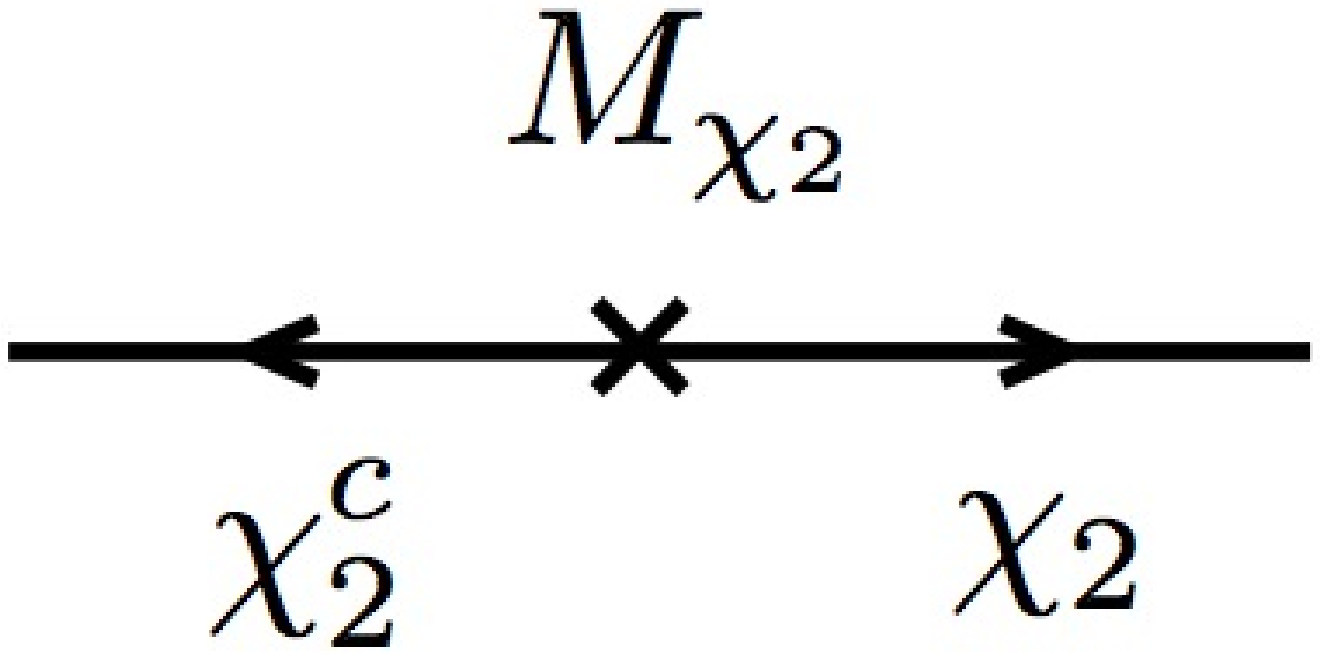}}
\lowpos{+}
\parbox{5cm}{\includegraphics[width=5cm]{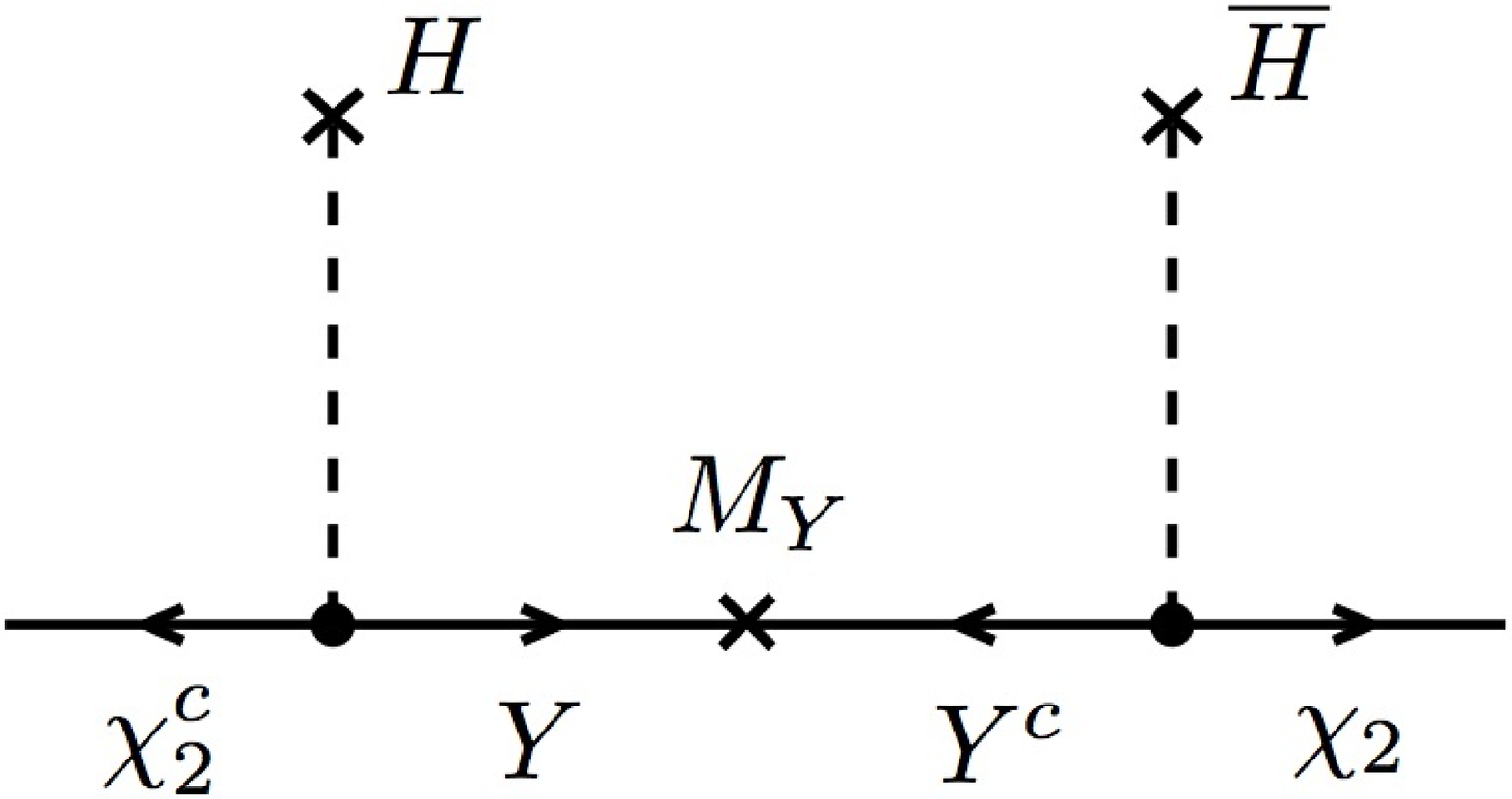}}
\lowpos{+ \ldots} \nonumber \eea
\caption{The large mass splitting in the
$\chi_{1}$ multiplet is
  achieved by means of a relatively suppressed singlet mass term
  accompanied by a particular set of ``level-2'' messengers giving
  rise to a mass term for $\chi_{1}^{u,\nu}$ only.
Since there is no need to split the masses of the
$\tilde{\chi}_{1}$
  multiplet the leading contribution to its mass can be dominated by the
  singlet mass term.
In principle, one can try to generate a mass splitting in the
  $\chi_{2}$ multiplet on similar grounds as for $\chi_{1}$; however,
  at the leading level the gauge structure admits only ``internal''
  contractions of the $H$ and $\overline{H}$ $SU(2)_{R}$ indices (and
  similarly, the $SU(2)_{L}$ indices of $\chi_{2}$ and $\chi_{2}^{c}$
  must saturate each other) and thus, regardless the quantum numbers
  of $Y$ the splitting in the $\chi_{2}$ sector is in general much
  milder.}
\label{messmass}
\end{figure}

This goal can be most economically achieved assuming that the
underlying dynamics of the $\chi_{1}$ field is governed by
interactions with an extra messenger $X$ that can propagate the
information about the Pati-Salam breaking (triggered by the VEVs
of $H$-fields) to the up-sector only. Indeed, this is possible if
the $SU(4)_{C}\otimes SU(2)_{L}\otimes SU(2)_{R}$ quantum numbers
of $X$ are chosen as $(15,1,1)$. In such a case the structure of
the $SU(2)_{R}$ contraction in the relevant operator
$\chi_{1}^{c}X \langle H\rangle$ picks up the component of  $H$
with nonzero VEV together with the up-type part of $\chi_{1}^{c}$
(i.e. $X$ corresponds to the diagonal matrix in the $2\otimes
\bar{2}$ product of $SU(2)_{R}$ and thus must be identified with
the singlet in $2\otimes \bar{2}=3\oplus 1$). The choice of $15$
out of $4\otimes\bar{4}=15\oplus 1$of $SU(4)_{C}$ is then
justified by the need to propagate the VEV not only to the
neutrino-like component $\chi_{1}^{\nu c}$ (as the singlet would
obviously do) but also to $\chi_{1}^{uc}$. This is illustrated
in Fig.\ref{messmass}.

%----

Notice that this mechanism is not applicable to the $\chi_{2}$
case as
 these messengers are $SU(2)_{R}$ singlets and the $SU(2)_{L}$
 symmetry prevents the up and down components from splitting up to the
 electroweak scale.
This is welcome as the  exchange of $\chi_{2}$ is assumed to be
the
 source of the universal 33 entries in the Yukawa sector. On the other
 hand, one can split the masses of its quark-like and lepton-like
 components $\chi_{2}^{q}$ and $\chi_{2}^{l}$, but this requires a
 messenger ($Y$) carrying both $SU(2)_{L}$ as well as $SU(2)_{R}$
 doublet indices. Moreover, to get $m_{\tau}$ slightly bigger than
 $m_{b}$ this calls for $M_{\chi_{2}^{l}}<M_{\chi_{2}^{q}}$ which can
 be achieved only if $Y$ is not
an $SU(4)_{C}$ singlet (otherwise the lepton part is picked up by
the
 VEVs of $H$ and $\bar{H}$). Thus, to achieve such a splitting at the
 lowest effective operator order one must make use of higher
 Pati-Salam representations that seem disfavoured by strings. Thus,
 sticking to small multiplets such an effect is expected to arise from
 higher order operators justifying the mildness of the GUT-scale
$b-\tau$ mass splitting.

\begin{table}[ht]\label{messengers2}
\centering
\begin{tabular}{|c|c|c|c|c|}
\hline
field & $SU(4)\otimes SU(2)_{L}\otimes SU(2)_{R}$ & $SO(3)$ & $U(1)$  & $Z_{2}$\\
\hline
$X$, $X^{c}$ &  $(15,1,1)$ & $3$ & $\mp 3$ & $ + $\\
$Y$, $Y^{c}$ &  $(15,2,2)$ & $1$ & 0 & $ - $ \\
\hline
\end{tabular}
\caption{ A sample ``level-2'' messenger sector giving rise to the
desired level-1 Dirac sector messenger mass splittings
$M_{\chi_{1}^{u}}\gg M_{\chi_{1}^{d}}$, $M_{\chi_{2}^{l}}\gtrsim
M_{\chi_{2}^{q}}$. }
\end{table}
Last note concerns the splitting in the $\tilde{\chi}_{1}$
messenger multiplet. We shall see in section \ref{numerics} that the numerical fit is perfectly
compatible with 
${M_{\tilde{\chi}_{1}^{u}}} \sim
{M_{\tilde{\chi}_{1}^{d}}} \sim {M_{\tilde{\chi}_{1}^{e}}}$ and thus there is
no need to generate a mass splitting within this sector. In other
words, we assume the masses of $\tilde{\chi}_{1}$ components to be
dominated by the explicit mass term in the superpotential.

%------------------------------------------
\subsection{The Majorana messenger sector}
%------------------------------------------
\label{majoranamesssector}
Concerning the Majorana mass terms given in \eqs{majlead}{majsublead}, we do not enter a
full analysis of the messenger sector here. The reason is that in
the Majorana case there is no need to adjust the messenger masses
in any particular way like in the Dirac sector, it is just enough
to ensure a mild hierarchy of the light Majorana masses, that
could be obtained in many different ways (c.f. the ``richness'' of
the set of subleading operators given by formula
(\ref{extraoperators1})).

\begin{figure}[ht]
\centering \bea
\parbox{3.5cm}{\includegraphics[width=3.5cm]{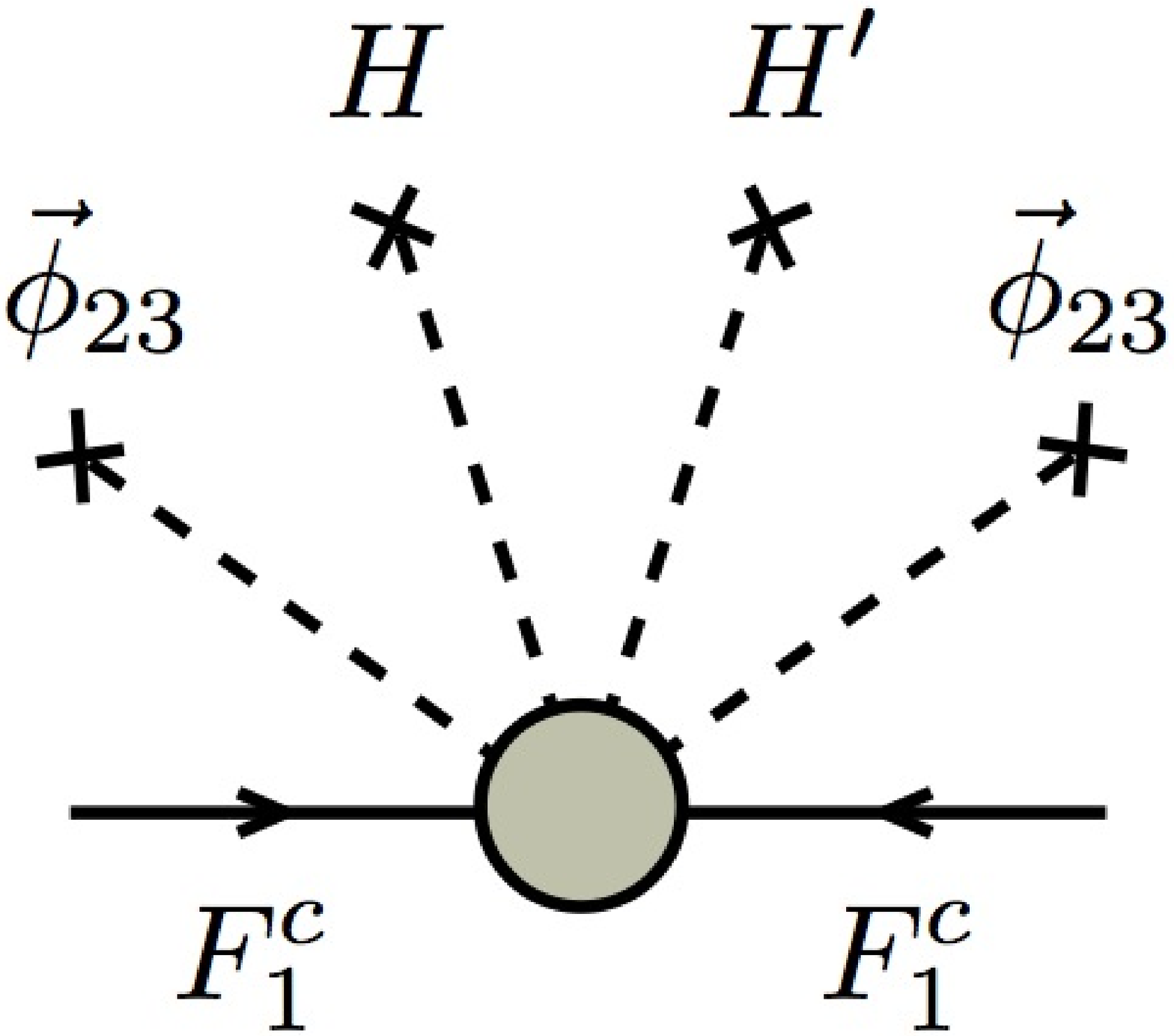}} &\lowpos{=}&
\parbox{5cm}{\includegraphics[width=5cm]{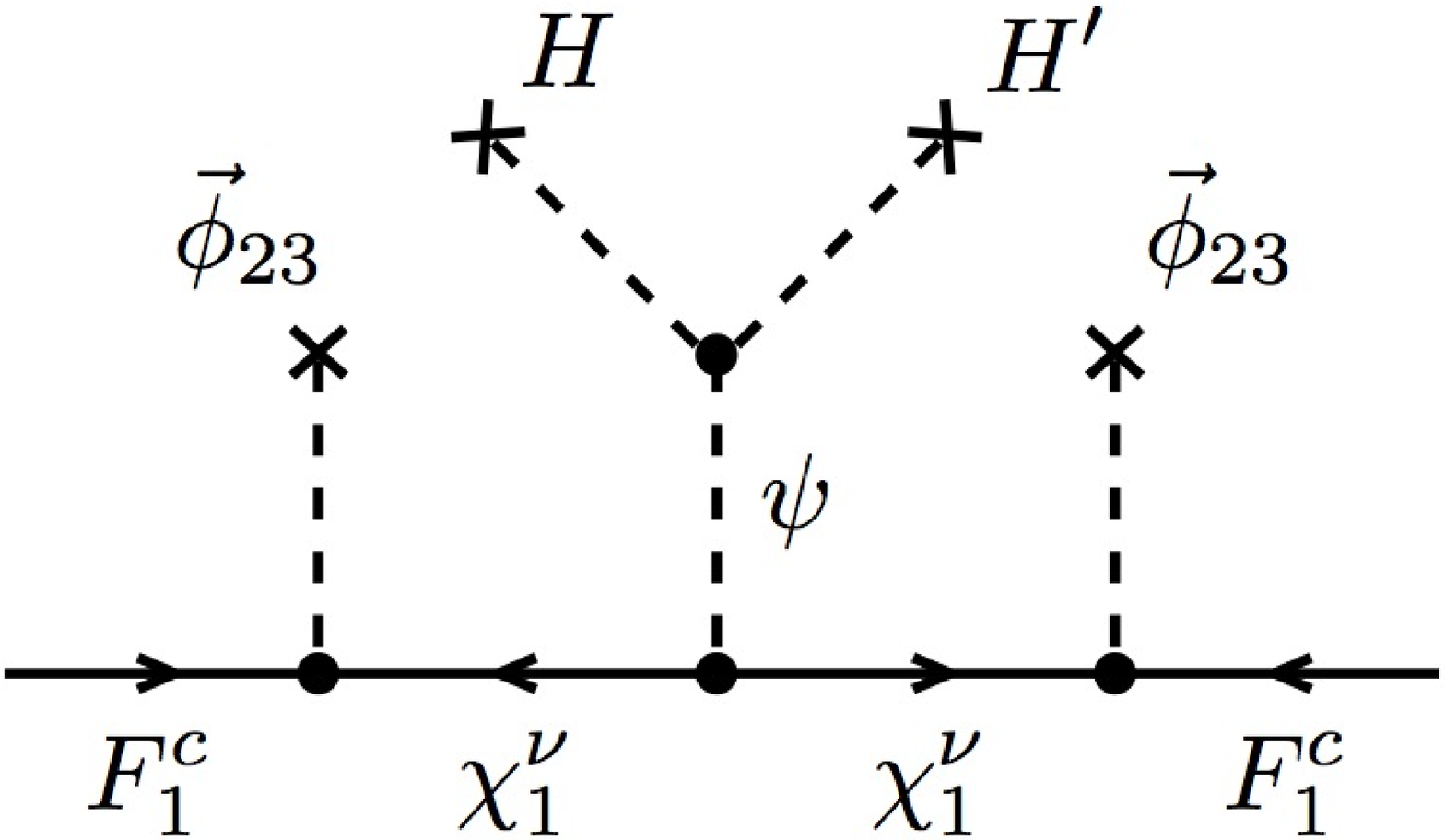}}\nonumber\\
\parbox{3.5cm}{\includegraphics[width=3.5cm]{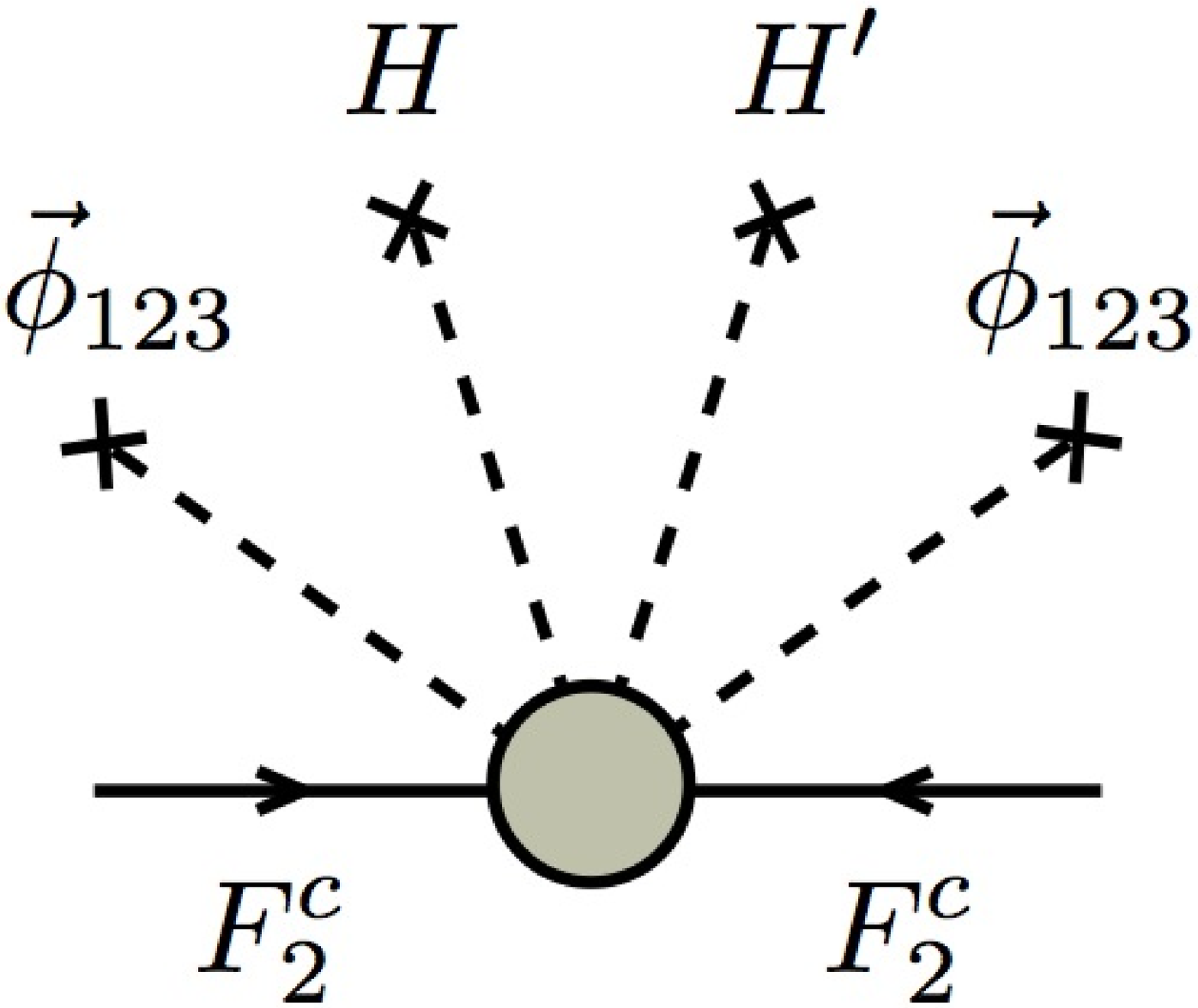}} &\lowpos{=}&
\parbox{5cm}{\includegraphics[width=5cm]{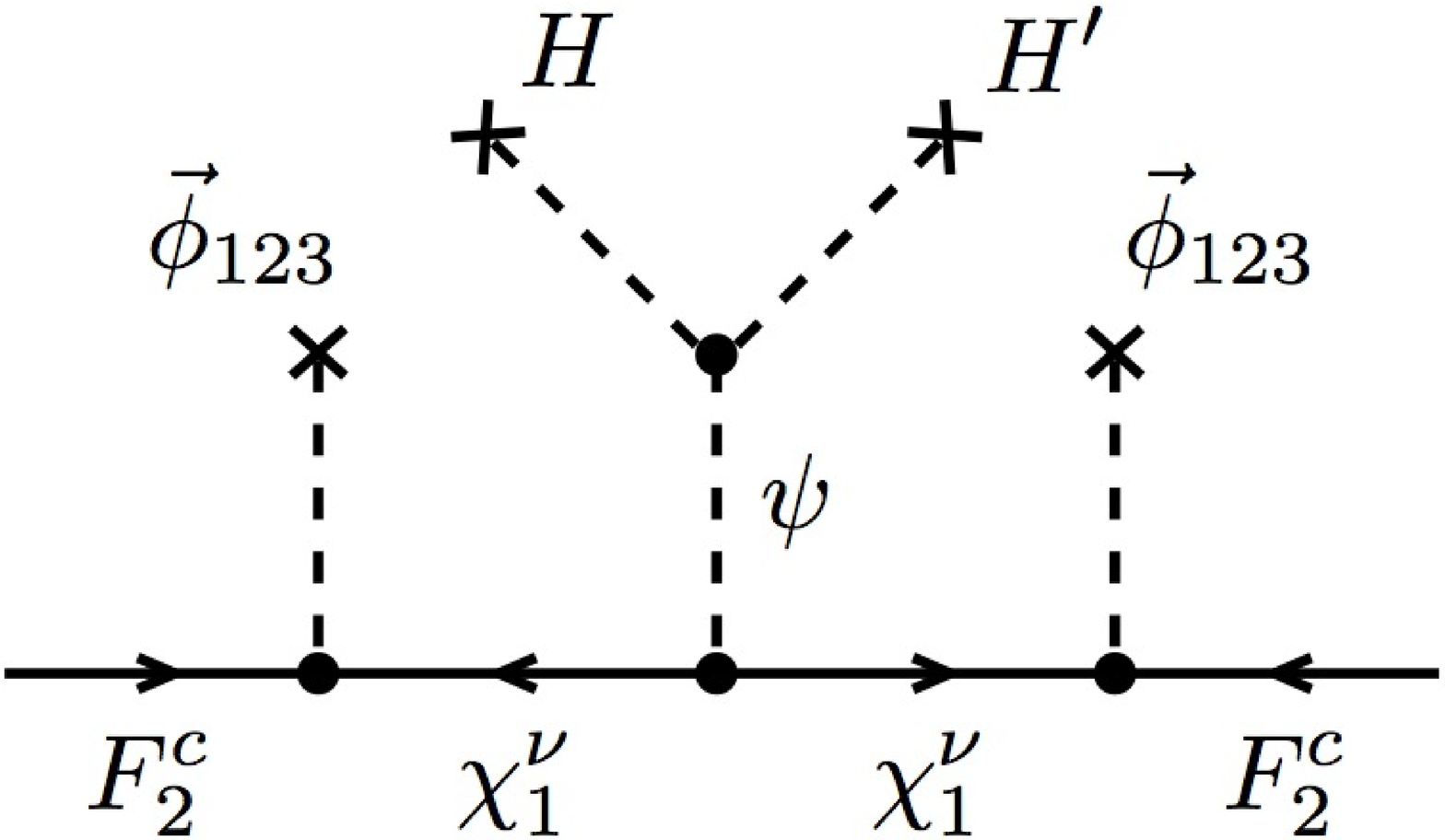}}\nonumber\\
\parbox{3.5cm}{\includegraphics[width=3.5cm]{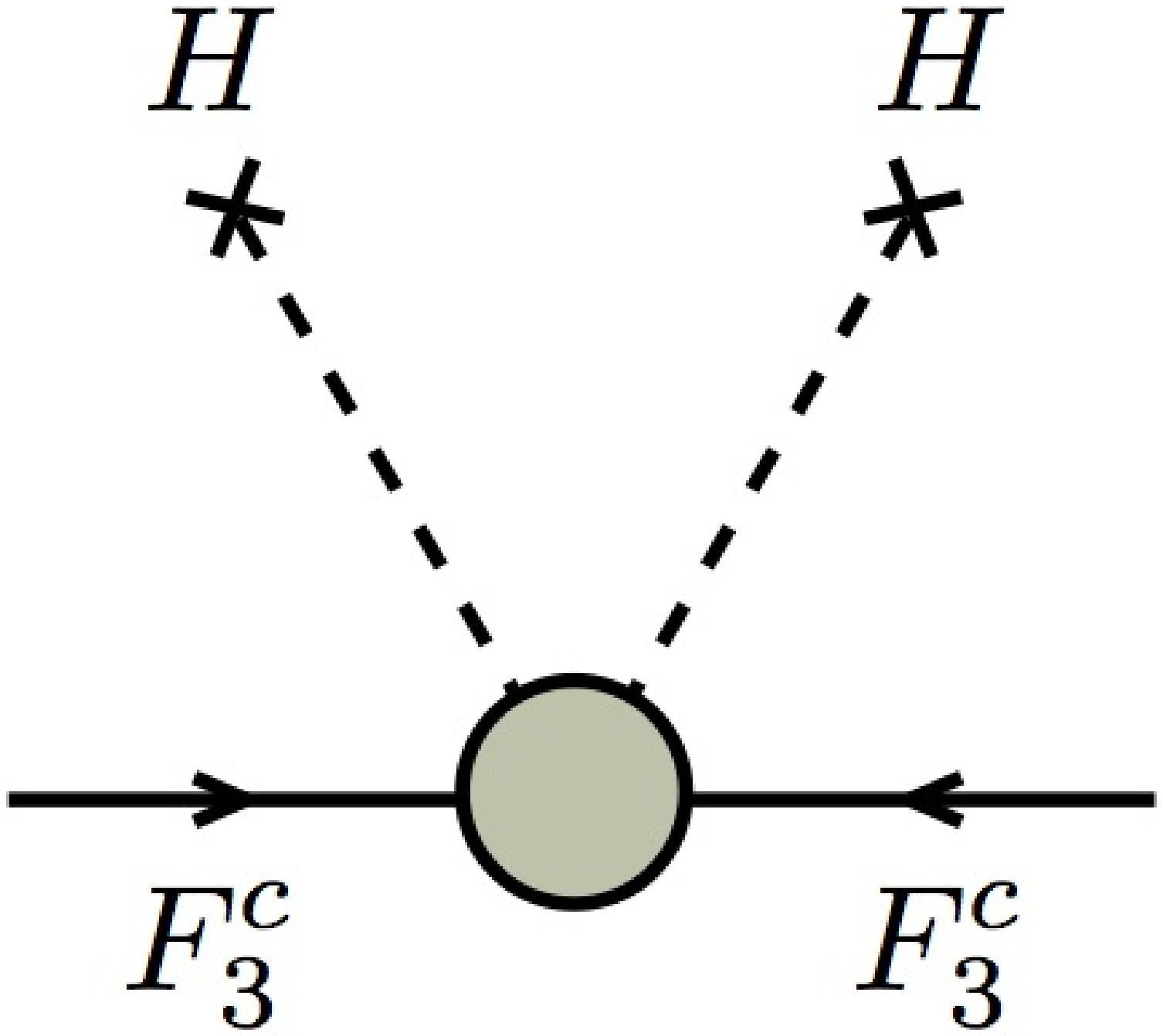}} &\lowpos{=}&
\parbox{5cm}{\includegraphics[width=5cm]{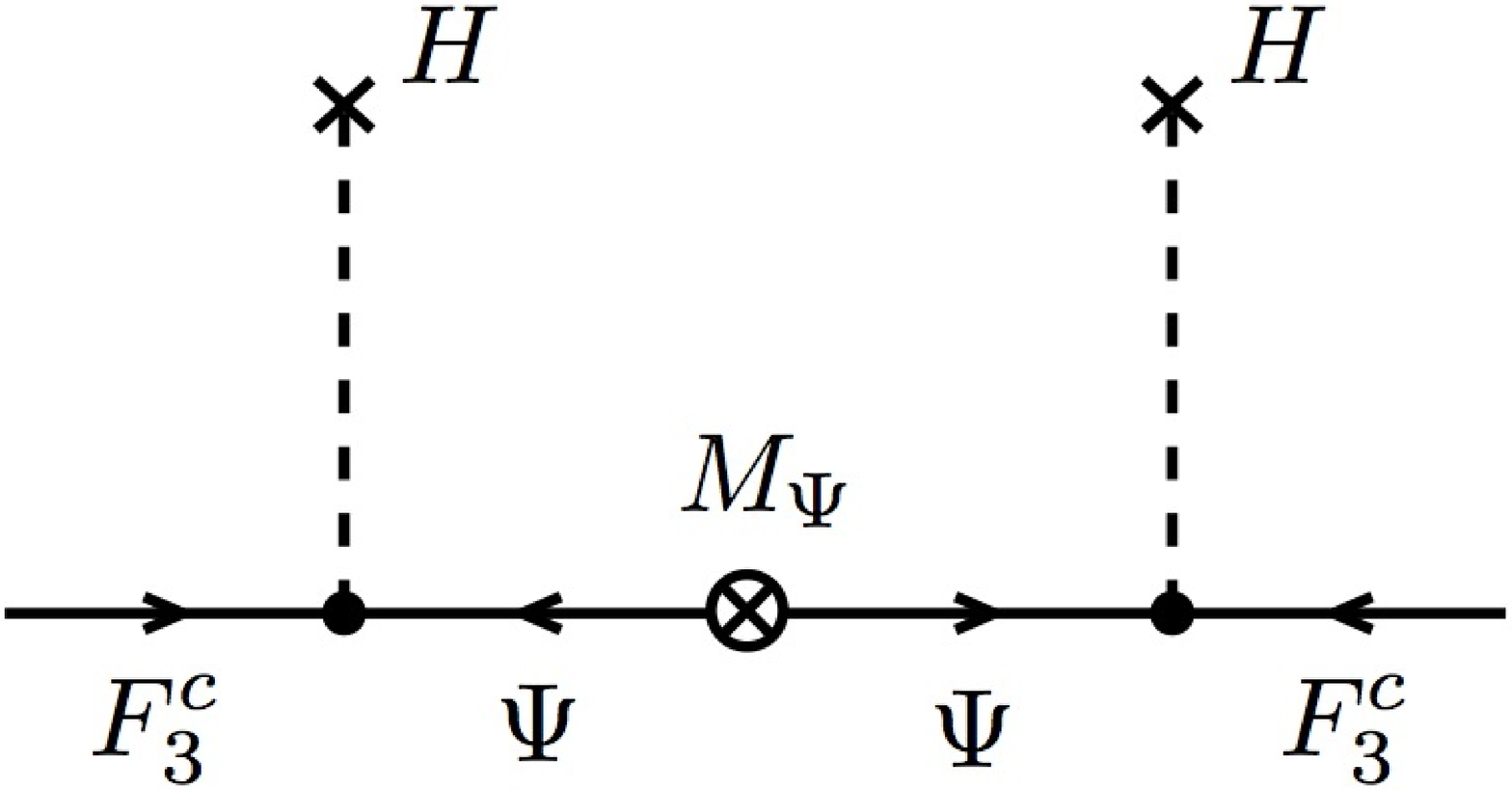}}\nonumber
\eea \caption{The basic lowest level (in number of extra
suppressions
  carried by the VEVs of $\phi_{23}$, $\phi_{123}$,$\ldots$ flavons ) effective operators contributing to the Majorana mass matrix in the neutrino sector. Note that there is no ``mixed'' term built up from the ``halves'' of the first and second graphs as the VEVs of $\phi_{23}$ and $\phi_{123}$ are orthogonal.}
\label{majorana-basic}
\end{figure}

We therefore restrict our discussion here to the universal leading operators
governing the diagonal entries of the mass matrix \footnote{The
choice of quantum numbers of the messenger $\psi$
  is  driven by simplicity, i.e. the need to pick up symmetric
  combinations from $4\otimes 4$ of $SU(4)_{C}$ and $2\otimes 2$ of
  $SU(2)_{R}$. Otherwise, the vertex with a pair of identical
  $\chi_{1}^{\nu}$ vanishes.}
(\ref{majoranapart2}). Their Froggatt-Nielsen structure is depicted
in Fig. \ref{majorana-basic}. We employ a pair of extra messengers
$\psi$ and $\Psi$ with the
  quantum numbers given in Table \ref{messengers1}.
Notice that the structure of these graphs is such that the Dirac
sector hierarchy associated with the $\vec{\phi}_{23}$ and
$\vec{\phi}_{123}$ VEVs is effectively cancelled in the seesaw
formula by the Majorana masses,
and thus the light neutrino mass splitting is
given by $m_2\sim m_3$ at lowest order.
This cancellation which was anticipated in section \ref{Majorana}
was based on the assumption of equal expansion parameters in the Dirac
and Majorana sectors. This assumption has now
been verified, since we have seen here that the explicit messenger sector
is common to both the Dirac and Majorana sector. Moreover, due to the presence of the heavy $\chi_{1}^{\nu}$ messenger obeying $M_{\chi_{1}^{\nu}}\sim M_{\chi_{1}^{u}}\gg M_{\chi_{1}^{d,e}}$ there is a further suppression in the effective values of the neutrino $\varepsilon_{x}^{\nu}$ factors as in the up-sector case, $\varepsilon_{x}^{\nu,u}\sim r \varepsilon_{x}^{e,d}$.
   
To ensure the proper splitting between the first and second
right-handed neutrino masses one can employ extra messenger fields
to construct similar diagrams for the other terms allowed by
symmetries, c.f. \eq{extraoperators1}.

%------------------------------------------
\section{Numerical analysis: a sample $\chi^{2}$ fit}
%------------------------------------------
\label{numerics}
At this point, all the ingredients are fully specified and we can
approach the fit of the quark masses and
CKM mixing parameters, using the charged lepton masses as inputs
\footnote{As we have already shown the
neutrino
  sector automatically leads to a tri-bimaximal mixing so we do not
  include the lepton mixing parameters into the $\chi^2$ analysis.}.

For the purposes of the numerical fit, it is convenient to
introduce the parameters
$y_{3}'\varepsilon_{3}^{q}=1$,
$a \delta \equiv y_{23}\varepsilon_{23}^{d}$, $b
\delta \equiv y_{123}\varepsilon_{123}^{d}$, $c \equiv
y_{GJ}\tilde{\varepsilon}_{23}^{d}\sigma^d$, $d \equiv
y_{3}\varepsilon_{3}^{d}$, $e \equiv
\tilde{y}_{23}(\tilde{\varepsilon}_{23}^{d})^{2}\varepsilon_{3}^{d}$
and  $f \delta \equiv
y_{12}\varepsilon_{12}^{d}\varepsilon_{3}^{d}$.
In terms of these parameters the charged sector
Yukawa matrices in \eqs{Yu}{Ye} can be
expressed as:
\bea Y^{d} & = &
\left(\begin{array}{ccc}
0 & b \delta & f \delta \\
a \delta & b \delta+c & e \\
-a \delta &b \delta+c & d+1
\end{array}\right) \\
Y^{u} & = & \left(\begin{array}{ccc}
0 & b \delta r & f \delta r^{2} \\
a \delta r & b \delta r - 2c r \tilde{r} & e r^{2} \tilde{r}\\
-a \delta r &b \delta r - 2c r \tilde{r} & d r +1
\end{array}\right) \\
Y^{e} & = & \left(\begin{array}{ccc}
0 & b \delta s & f \delta s^{2} \\
a \delta s & b \delta s +3 c s \tilde{s} & e s^{2} \tilde{s}\\
-a \delta s &b \delta s + 3 c s \tilde{s} & d s + t
\end{array}\right)
\eea 
where 
\be r\equiv \frac{M_{\chi_{1}^{d}}}{M_{\chi_{1}^{u}}},
\quad \tilde{r}\equiv
\frac{M_{\tilde{\chi}_{1}^{d}}}{M_{\tilde{\chi}_{1}^{u}}}, \quad
s\equiv \frac{M_{\chi_{1}^{d}}}{M_{\chi_{1}^{e}}}, \quad
\tilde{s}\equiv
\frac{M_{\tilde{\chi}_{1}^{d}}}{M_{\tilde{\chi}_{1}^{e}}}, \quad t
\equiv \frac{M_{\chi_{2}^{q}}}{M_{\chi_{2}^{l}}}. 
\ee

As an example of a successful fit,
we shall present here a sample set of values of the parameters defined
above leading to a very good agreement with the experimental data
evolved (using the MSSM Yukawa running \cite{Das:2000uk}) up to the GUT-scale,
which loosely corresponds to the textures in Eq.\ref{Roberts},
although here of course they originate from a dynamical model.
Notice that all the unsuppressed parameters (i.e. everybody up to
$\varepsilon^{d}_{23}$, $\varepsilon^{d}_{123}$ and
$\varepsilon^{d}_{12}$) fall in the natural ${\cal O}(1)$ domain.
Next, the fit is obtained under the assumption that {\it there is
only a single parameter $\delta$
  governing all the ad-hoc suppressed quantities}
$\varepsilon^{d}_{23}$, $\varepsilon^{d}_{123}$ and
$\varepsilon^{d}_{12}$, i.e. $\varepsilon^{d}_{23,123,12}\propto
{\cal
  O}(1) \times \delta $.
In accord with the messenger sector dynamics we let $r$ (c.f.
section \ref{messenger-spectra}) depart from 1 (we have already
seen that its preferred value in order to reproduce the textures
(\ref{Roberts}) is roughly $r \sim 0.04$) but keep all the other
messenger mass ratios around $1$.

To obtain the desired structure in Eq.\ref{Roberts} (assuming
$y_{3}'\varepsilon_{3}^{q}\sim 1$ and keeping all the Yukawa
couplings at ${\cal O}(1)$ level) one can estimate the magnitude
of the extra suppression factors $\delta_{23,123}$ in
$\varepsilon^{d}_{23}$ and $\varepsilon^{d}_{123}$ to be about \be
\varepsilon^{d}_{23} \sim \varepsilon^{d}_{123}\sim 0.003 \ee To
make this suppression potentially natural (i.e. universal), also
the small extra factor in the VEV of $\phi_{12}$ should have
roughly the same value and we should assume
$\varepsilon^{d}_{12}\sim 0.003$ as well. Is this compatible with
the physical value of the 1-3 CKM mixing angle ? Notice that with
$r\sim 0.04$ the 13 term in the up Yukawa matrix is strongly
suppressed and $\theta_{13}^{q}$ comes entirely from the down
sector. Thus, $\theta_{13}^{q}\sim 0.003$ is obtained in a
completely  natural way provided $|y_{12}\varepsilon_{3}^{d}|\sim
1$, that corresponds to our assumption of no extra suppression
factor in the $\phi_{3}$ VEV.

It can be easily checked that a good fit of all the quark and
charged lepton data %(barring few-$\sigma$ deviations for the first generation masses that shall be anyway affected upon passing beyond the leading order approximation) 
can be obtained for
instance for the values of the relevant parameters given in Table
\ref{Tablechisq}.

\begin{table}[ht]
\centering
  \begin{tabular}{|c|r|}
  \hline
  observable & input data \\
  \hline
  $m_u$ [MeV]   & $0.55 \pm 0.25$\\
  $m_c$ [MeV]   & $210 \pm 0.21$\\   % this is the error used in the fit
  $m_t$ [GeV]   & $82.4^{+30.3}_{-14.8}$\\
  $m_d$ [MeV]   & $1.24 \pm 0.41$\\
  $m_s$ [MeV]   & $21.7 \pm 5.2$\\
  $m_b$ [GeV]   & $1.06^{+0.14}_{-0.09}$\\
  $m_e$ [MeV]   & $0.358$\\
  $m_\mu$ [MeV]   & $75.67$\\
  $m_\tau$ [GeV]   & $1.292$\\
  $\sin\phi^\mathrm{CKM}_{12}$ & $0.2243\pm 0.0016$\\
  $\sin\phi^\mathrm{CKM}_{23}$ & $0.0351\pm 0.0013$\\
  $\sin\phi^\mathrm{CKM}_{13}$ & $0.0032\pm 0.0005$\\
  $\delta_\mathrm{CKM}$        & $60^\circ \pm 14^\circ$\\
  \hline
  \end{tabular}
\,\,\,\,\,\,\,\,\,\,\,\,\,\,\,\,\,\,\,\,\,\,
\begin{tabular}{|c|r|r|}
\hline
parameter & value (case 1)&  value (case 2)\\
\hline
$\delta$ & $0.003$  & $0.003$\\
\hline
$a$ & $2.100 $ & $1.657 $\\
$b$ & $1.240 e^{1.488i} $ & $1.220 e^{1.565i} $ \\
$c$ & $0.026  e^{4.667i}  $ & $0.026  e^{4.730i}  $\\
$d$ & $1.017 e^{2.262i}$  & $0.913 e^{2.262i}$  \\
$e$ & $0.032  e^{0.223i} $ & $0.030  e^{0.168i} $ \\
$f$ & $1.059  e^{1.304i} $  & $0.892  e^{1.350i} $ \\
\hline
$r$ & 0.040 & 0.040\\
$\tilde{r}$ & 1 & 1\\
$s$ & 0.773  & 0.773\\
$\tilde{s}$ & 1 & 1 \\
$t$ & 1.252  & 1.252 \\
\hline
\end{tabular}
\\
\mbox{}
\\
\begin{tabular}{|c|r|r|r|r|}
  \hline
 $\chi^2$ & \multicolumn{2}{c|}{case 1} &  \multicolumn{2}{c|}{case 2} \\
\hline
  observable & prediction & pull ($\sigma$) & prediction & pull ($\sigma$)\\
  \hline
  $m_u$ [MeV]   & $1.54$ & $+6.33$  & $0.96$ & $+1.82$ \\
  $m_c$ [MeV]   & $207.8$ & $-0.72$ & $209.0$ & $-0.13$\\
  $m_t$ [GeV]   & $90.13$ & $+0.01 $ & $90.85$ & $+0.04 $\\
  $m_d$ [MeV]   & $1.45$ & $-0.50$  & $1.12$ & $-1.29$ \\
  $m_s$ [MeV]   & $30.74$ & $+1.74$ & $30.30$ & $+1.66$\\
  $m_b$ [GeV]   & $1.15$ & $+0.54 $  & $1.09$ & $ 0.00 $ \\
  $m_e$ [MeV]   & $0.358$ & $\times$  & $0.358 $ & $\times$ \\
  $m_\mu$ [MeV]   & $75.67$ &  $\times $ & $75.67$ & $\times$ \\
  $m_\tau$ [GeV]   & $1.292$ & $ \times $  & $1.292$ & $\times$ \\
  $\sin\phi^\mathrm{CKM}_{12}$ & $0.2231$ & $ +0.09 $ & $0.2227$ & $ -0.10 $\\
  $\sin\phi^\mathrm{CKM}_{23}$ & $0.0372$ & $ +0.35 $ & $0.0370$ & $ +0.21 $\\
  $\sin\phi^\mathrm{CKM}_{13}$ & $0.0033$ & $ +0.17 $ & $0.0032$ & $ 0.00 $\\
  $\delta_\mathrm{CKM}$   & $73.9^{o}$ & $ +1.00 $ & $65.7^{o}$ & $ +0.41 $\\
  \hline
  Total: & $\chi^{2}$ & $45.37$ & $\chi^{2}$ & $7.95$\\
  \hline
\end{tabular}
\caption{\label{Tablechisq} A simple $\chi^{2}$ fit of the quark
and charged lepton masses and the CKM mixing parameters. The
GUT-scale input data are taken from \cite{Das:2000uk} in view of
the slight update advocated in \cite{Bertolini:2006pe} and
references therein. Notice the irrelevance of the phase of $a$
which affects only the first columns of the relevant Yukawa
matrices and thus can be rotated away. In case 1 the analysis is
performed at the leading order in the number of flavon insertions,
which leads to small deviations in the
first generation masses governing the total $\chi^{2}$
function. The second column (case 2) gives an example of a very good fit in case
of a
tiny (but nonzero) 11 entry potentially arising at next-to-leading
order; for details see the text.
}
\end{table}
The solution called ``case 1'' corresponds to our best fit of the
leading
Yukawa structures given above and leads to the
following hierarchies at the high energy scale:
\bea |Y^{d}| & \doteq & \left(\begin{array}{ccc}
0 & 0.0037 & 0.0031 \\
0.0063 & 0.02228 & 0.0317 \\
0.0063 & 0.02228 & 0.8586
\end{array}\right) \\
|Y^{u}| & \doteq & \left(\begin{array}{ccc}
0 & 0.0001 & 0 \\
0.0003 & 0.0023 & 0.0001 \\
0.0003 & 0.0023 & 0.9745
\end{array}\right)\\
|Y^{e}| & \doteq & \left(\begin{array}{ccc}
0 & 0.0028 & 0.0019 \\
0.0049 & 0.0574 & 0.0189 \\
0.0049 & 0.0574 & 0.9644
\end{array}\right)
\eea
One can verify easily that the
portion of the lepton mixing coming from the charged lepton Yukawa
is negligible. Moreover, the Cabibbo mixing emerges predominantly
from the down-type Yukawa \`{a} la Gatto et al. \cite{Gatto:1968ss} which
is welcome. Next, there seems to be preference of $s$ slightly below
$1$ leading to
$m_s$ in the upper $\sigma$ region. This seems to originate from the
fact that the optimal $M_{\chi^d_1}/M_{\chi^u_1}$ ratio $\sim 0.04$
leads to a natural value for the double ratio of the physical masses
\be
x^{fit}=\frac{m_c/m_t}{m_s/m_b}\sim 2r\sim 0.08
\ee
and thus $m_s$ should be in its upper-$\sigma$ region to lower the
central value $x^c\sim 0.116$. Consequently, the pure Georgi-Jarlskog
relation $m_s/m_\mu=1/3$ (perfectly valid for $m_s$ around 25 MeV
around the GUT scale) is slightly violated. % and $s$ must be below 1.
Note that since $s$ parametrizes the mass splitting of the type-1
messengers, their masses are anyway expected to differ after the
Pati-Salam symmetry breaking and there is no technical problem to
receive $s$ around $0.8$ as suggested by the numerics.

Remarkably enough, around $90\%$ of the total $\chi^2$ comes from
the first generation masses that are quite sensitive to the subleading
corrections. Indeed, even as tiny as order $10^{-5}$ corrections
to the relevant Yukawa
matrices allow for a dramatic improvement of the $\chi^2$ value,
c.f. Table  \ref{Tablechisq}, ``case 2''\footnote{In the present case
  (``case 2'' in Table \ref{Tablechisq}) we
  allowed for a variation of order $10^{-5}$ at the 11 positions of the
 (down-type) Yukawa entries under consideration (corresponding typically to a
 negligible factors like $10^{-7}$ emerging in the up-type ones).}. Recall that such extra
factors emerge in a natural way from higher order effective operators like
for instance
\be
W^{h.o.}= y_{a}\frac{1}{M^2}\vec{F}.({\vec{\phi}}_{23}\times
\vec{\phi}_{12})F_1^c h+ y_{b} \frac{1}{M^3}\vec{F}.(\tilde{\vec{\phi}}_{23}\times
\vec{\phi}_{12})F_2^c\Sigma h +  \ldots
\ee
Moreover, since $\phi_{12}$ is a $U(1)\otimes Z_2$ singlet such
terms emerge in a natural way upon inserting the $\phi_{12}$ VEV
into the existing operators (\ref{diracsuperpot1}) without any need to enlarge the messenger sector.

Although the charged lepton masses and mixing angles are not precisely
of the Georgi-Jarlskog type they do have the
same qualitative form, thus for example 
$\theta_{12}^e\approx 0.05 \sim \theta_{12}^d/3$.
This means that the charged lepton corrections to 
tri-bimaximal neutrino mixing cannot be precisely
related to the Cabibbo angle.
Nevertheless we find the physical lepton mixing angles
\cite{King:2005bj,Antusch:2005kw}:
\be
\theta_{13}\approx 2^o
\ee
\be
\theta_{12}+\theta_{13}\cos (\delta_{MNS} - \pi )\approx 35.26^o
\label{sumrule}
\ee
where Eq.\ref{sumrule} is the sum rule \cite{King:2005bj,Antusch:2005kw}
where $\delta_{MNS} $ is the MNS CP phase which enters neutrino oscillations,
and $35.26^o$ follows from tri-bimaximal neutrino mixing.

With the fit parameters in hand, we see that
the consistency of the model requires the following hierarchy of
the various mass scales present:
\be
M_{X,Y}>M_{GUT}>M_{\chi_{1}^{u}}\gg M_{\tilde{\chi}_{1}}\gtrsim
M_{\chi_{1}^{d}}\gtrsim M_{\chi_{2}^{q}}\sim M_{\chi_{2}^{l}}>
\langle \Sigma \rangle, \langle \phi_{3} \rangle, \langle
\tilde{\phi}_{23} \rangle \gg \delta  \langle \phi_{23} \rangle,
\delta \langle \phi_{123} \rangle,  \delta \langle \phi_{12}
\rangle.
\ee
The first two inequalities satisfy the need to
generate the proper effective masses of the Dirac sector messenger
fields dynamically via exchange of the ``level-2'' messengers $X$
and $Y$. The third, fourth, fifth and sixth relation ensures the
proper Dirac sector messenger hierarchies described above. The
last two relations justify the expansions in the number of
$\Sigma$ and flavon insertions used throughout the analysis. It is
obvious from the $\chi^{2}$ parameters that there is no problem to
satisfy these relations with all the relevant tree-level Yukawa
couplings (c.f. \eq{couplings}) at the ${\cal O}(1)$ level. Thus,
the presented fit is natural.
Notice also that the mildly suppressed 22, 32 and 23 Yukawa
entries are reasonable as they arise from higher order effective
operators and thus for instance $\tilde{\varepsilon}_{23}^{d}\sim
0.2$ (as suggested by the fit).
All parameters are therefore either of order unity, or
their smallness is accounted for by a ratio of mass scales,
apart from the so far unexplained smallness of the parameter $\delta$.
In the next section we shall interpret $\delta$
as a volume suppression factor emerging in a higher dimensional theory.

%---------------------------
\section{The 5d orbifold GUT model based on $SO(3)\times SO(10)$ }
%---------------------------
\label{5dmodel}
%---------------------------
%%%%%%%%%%%%%%%%%%%%%%%%%%%%%%%%%%%%%%%%%%%
\subsection{Introduction}
%%%%%%%%%%%%%%%%%%%%%%%%%%%%%%%%%%%%%%%%%%%

The ad-hoc suppression factors present in the 4d model we proposed
in the first part of this paper receive a simple justification once
 the theory is promoted to more than 4 dimesions. Indeed, every field
 propagating in the higherdimesional bulk receives (from the point of
 view of the 4d effective theory) a volume suppression factor that can
 be used to generate extra hierarchies in its couplings to the
 localized fields.

As we have shown in section \ref{numerics}, the physical observables
can be fitted even under the nontrivial assumption that the extra suppression
factors in the effective couplings of the VEVs of the $\phi_{23}$,
$\phi_{123}$, $\phi_{12}$ flavons (and an extra  Higgs pair) all coincide. Thus, it
is natural to ask whether such an effective 4d model can be understood as
a low-energy limit of a more fundamental 5d theory.

Moreover, unlike the 4d Pati-Salam model, such an embedding could be
viewed as a ``true'' 5d grand unified model that (as an orbifold GUT)
can naturally accommodate the incomplete GUT multiplets of the
effective model provided they live on the orbifold fixed point with a
reduced gauge symmetry.

%%%%%%%%%%%%%%%%%%%%%%%%%%%%%%%%%%%%%%%%%%%
\subsection{The setup}
\begin{figure}[ht]
\centering
\includegraphics[width=0.45\textwidth]{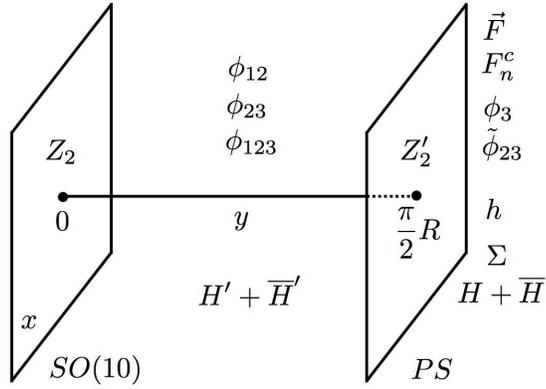}
  \caption{The physical 5-d setup: $S^{1}/Z_{2}\times Z'_{2}$ orbifold is used to break the 5-dimensional N=1 SUSY in the bulk to the 4-d N=1 SUSY of the MSSM, and provides for the first step in the symmetry breaking of the unified $SO(10)$ gauge group. The geometrical location of some of the flavon fields yields the necessary suppression factors in the relevant Yukawa textures.}
\label{orbifold}
\end{figure}

%%%%%%%%%%%%%%%%%%%%%%%%%%%%%%%%%%%%%%%%%%%
We assume a variation of the ``standard'' 5-dimensional $SO(10)$ SUSY
GUT compactified on the $S^{1}/Z_{2}\times Z'_{2}$ orbifold
\cite{Dermisek:2001hp, Kim:2002im, Alciati:2006sw,Kim:2004vk}. The
first $Z_{2}$ orbifold
 projection acting on the fifth coordinate $y$ as $y\to -y$ is used
to reduce the 5-dimensional N=1 supersymmetry (equivalent to N=2 SUSY
in 4 dimensions) to an effective 4-dimensional N=1 SUSY while the full
$SO(10)$ gauge symmetry is reduced to the Pati-Salam $SU(4)\otimes
SU(2)_{L}\otimes SU(2)_{R}$ on the brane located at the fixed point
(the Pati-Salam brane) of the second projection $Z'_{2}: y\to \pi R
-y$ . In accordance to the 4d model we  assume an $SO(3)$ flavor
symmetry acting on the left-handed matter multiplets living as
incomplete SO(10) multiplets at the Pati-Salam brane. As before, the
flavor symmetry is augmented by an aditional global $U(1)\otimes Z_{2}$.
The graphical representation of the setup is given in
Fig. \ref{orbifold}.

 The full set of Pati-Salam brane fields used in our construction can
 be found in
Table \ref{tab-brane}.
We take the advantage of the reduced gauge symmetry to put all the
 relevant Pati-Salam $SU(4)\otimes
SU(2)_{L}\otimes SU(2)_{R}$ matter multiplets here.
In order to get ${\cal O}(1)$ Yukawa
couplings for the third generation and only a mild  suppression of the
 second generation masses the Georgi-Jarlskog Higgs field
 $\Sigma$ as well as the light Higgs bidoublet live there\footnote{Notice that there
is no doublet-triplet splitting problem associated to the light
doublets since they enter the game as an incomplete $SO(10)$
multiplet.} together with the flavons that should couple to the matter
 bilinears without further suppression factors ($\phi_3$ and $\tilde{\phi}_{23}$).
\begin{table}[ht]
\centering
\begin{tabular}{|c|c|c|c|c|c|}
\hline
field & $SO(10)$ (incomplete) & $SU(4)\otimes SU(2)_{L}\otimes SU(2)_{R}$ & $SO(3)$ & $U(1)$  & $Z_{2}$\\
\hline
$\vec{F}$ & $16$ & $(4,2,1)$ & $3$ & 0 & $+$\\
$F^{c}_{1}$ & $16$   & $(\overline{4},1,2)$ & $1$ & $+2$ & $-$ \\
$F^{c}_{2}$  & $16$ & $(\overline{4},1,2)$ & $1$ & $+1$ & $+$\\
$F^{c}_{3}$   & $16$ & $(\overline{4},1,2)$ & $1$ & $-3$ & $-$\\
\hline
$h$  & $10$ & $(1,2,2)$ &  $1$ & 0 & $+$\\
$H$, $\overline{H}$  & $16\oplus \overline{16}$ & $(4,1,2)$, $(\overline{4},1,2)$ & $1$ & $\pm 3$ & $+$
\\
\hline
$\vec{\phi}_{3}$  & $1$ & $(1,1,1)$ & $3$ & $+3$ & $-$ \\
\hline
$\Sigma$  & $210$ & $(15,1,3)$  &$ 1 $ & -1 & $-$\\
$\vec{\tilde{\phi}}_{23}$ & $1$ & $(1,1,1)$ & $3$ & $0$ & $-$ \\
\hline
\end{tabular}
\caption{\label{tab-brane} The brane fields:
the basic Higgs, matter and flavon content of the model.
}
\end{table}

To obtain a natural suppression of the first and second family Yukawa
couplings we put
the flavons responsible for these interactions to the bulk,
i.e. $\phi_{23}$, $\phi_{123}$ and $\phi_{12}$ are assumed to propagate in 5
dimensions. Since there is a need to arrange an extra suppression
factor in the first and second generation Majorana masses the relevant
``primed'' Higgs pair ($H'\oplus \overline{H'}$) also lives in the bulk.

\begin{table}[ht]
\centering
\begin{tabular}{|c|c|c|c|c|c|}
\hline
field & $SO(10)$ & $SU(4)\otimes SU(2)_{L}\otimes SU(2)_{R}$ decomposition & $SO(3)$  & $U(1)$ & $Z_{2}$\\
\hline
$H'$, $\overline{H}'$  & $16\oplus \overline{16}$ & $(\overline{4},1,2)$, $(\overline{4},2,1)$ & $1$ & $\mp 3$ & $+$
\\
\hline
$\vec{\phi}_{23}$  & $1$ & $(1,1,1)$ & $3$ & $-2$ & $-$\\
$\vec{\phi}_{123}$  & $1$ & $(1,1,1)$ & $3$ & $-1$ & $+$\\
\hline
$\vec{\phi}_{12}$ & $1$ & $(1,1,1)$ & $3$ & $0$ & $+$\\
\hline
\end{tabular}
\caption{\label{bulkfields}
The bulk fields:
due to the bulk location of these multiplets
the effective 4-d VEVs of the fields coupled to
the Pati-Salam brane matter multiplets are ``screened''
by the bulk suppression factors entering the relevant 4-d vertices.
}
\end{table}

As we shall see, such a setup upon compactification naturally leads to the
effective Pati-Salam model constructed in previos sections with the desired Yukawa textures for the
Dirac and Majorana mass matrices.

%%%%%%%%%%%%%%%%%%%%%%%%%%%%%%%%%%%%%%%%%%%
\subsection{The 5d superpotential}
%%%%%%%%%%%%%%%%%%%%%%%%%%%%%%%%%%%%%%%%%%%
The construction goes
along similar lines as in the 4d case. In 5d, the relevant pieces of
superpotential read:
\be\label{Dirac}
W_{Y}^{leading}\propto\int {\rm d}y\, \delta\left(y-\frac{\pi}{2}R\right)
\frac{1}{M_{m}}\left(
\frac{1}{\sqrt{M_{*}}}
y_{23}\vec{F}.\vec{\phi_{23}}F_{1}^{c}+
\frac{1}{\sqrt{M_{*}}}
y_{123}\vec{F}.\vec{\phi_{123}}F_{2}^{c}+
y_{3}\vec{F}.\vec{\phi_{3}}F_{3}^{c}
\right)h
\ee
\be
W_{Y}^{subl.}
\propto \int {\rm d}y\, \delta\left(y-\frac{\pi}{2}R\right)
\times
\ee
$$
\times\left[
\frac{1}{M_{m}^{3}}
\tilde{y}_{23}\vec{F}.\vec{\tilde{\phi}}_{23}(\vec{\phi}_{3}.\vec{\tilde{\phi}}_{23}) F_{3}^{c}h+
\frac{1}{M_{m}^{2}\sqrt{M_{*}}}
y_{12}\vec{F}. (\vec{\phi}_{3}\times \vec{{\phi}}_{12}) F_{3}^{c}h
+\frac{1}{M_{m}^2}
y_{GJ}\vec{F}.\vec{\tilde{\phi}}_{23}F_{2}^{c}h \Sigma \right] +\ldots
$$
\be
W_{M}^{leading}
\propto \int {\rm d}y\, \delta\left(y-\frac{\pi}{2}R\right)
\left[
\frac{1}{M_{m}}w_{3}{F_{3}^{c}}^{2}H^{2}
+\frac{1}{{M_{*}}^{3/2}M_{m}^{3}}
\left(
w_{1}{F_{1}^{c}}^{2}H H' \phi_{23}^{2}+
w_{2}{F_{2}^{c}}^{2}H H' \phi_{123}^{2}
\right)
\right]
\ee
$$
W_{M}^{subl.}
\propto \int {\rm d}y\, \delta\left(y-\frac{\pi}{2}R\right)
\left[
\frac{1}{M_{m}^{4}M_{*}^{3/2}}{F_{2}^{c}}^{2}H H' (\vec\phi_{23} \times \vec\phi_{12}).\vec{\tilde\phi}_{23}
+
\frac{1}{M_{m}^{4}M_{*}^{2}}{F_{1}^{c}}^{2}H'H' (\vec\phi_{3} \times \vec\phi_{123}).\vec{\tilde\phi}_{23}+
 \right.
$$
\be
\left.+\frac{1}{M_{m}^{4}M_{*}^{3/2}}{F_{1}^{c}}{F_{3}^{c}}
H H' (\vec\phi_{3} \times \vec\phi_{23}).\vec{\phi}_{12}+
\frac{1}{M^{4}M_{*}^{2}}{F_{1}^{c}}{F_{2}^{c}}
\right]+\ldots
\ee
where as before $M_{m}$ stands for the masses of the Froggatt-Nielsen
messenger fields, $R$ corresponds to the volume of the extra dimension and the dimensionfull
parameter $\sqrt{M_{*}}$ is associated with every field propagating in
the bulk to keep the action dimensionless.
It is then easy to see that integrating over $y$ and assigning
\be
\delta\equiv \frac{1}{\sqrt{2\pi R M_*}}
\ee
we recover all the relevant 4d operators of section \ref{4dsection} with a natural
bulk-suppression $\delta$ in all the factors including VEVs of
$\phi_{23}$,$\phi_{123}$,  $\phi_{12}$ and $H'$.

Thus, the 4d theory can be viewed as a low-energy limit of a 5d
orbifold GUT model.

%---------------------------
\section{Conclusions}
%---------------------------
The problem of fermion masses and mixings has become more interesting
over recent years with the discovery of neutrino mass and mixings,
which show that the neutrino sector differs markedly from the
charged fermion sector. The most promising approaches to 
understanding fermion masses and mixings seem to involve a
combination of GUT and Family Symmetries.
The most recent neutrino oscillation data
is consistent with tri-bimaximal mixing, which
could naturally result from the see-saw mechanism
with CSD where a non-Abelian
Family Symmetry such as $SU(3)$, $SO(3)$,
or one of its discrete subgroups, provides
a framework for the necessary vacuum alignment of flavon VEVs
\cite{King:2005bj,deMedeirosVarzielas:2005ax,deMedeirosVarzielas:2006fc,deMedeirosVarzielas:2005qg}.
We have considered a specific 4d model  
based on Pati-Salam unification and $SO(3)$ gauged Family Symmetry,
although it could be extended to the case where $SO(3)$
is replaced by one of its discrete subgroups
such as $A_4$, where the problem of vacuum
alignment is potentially simpler \cite{deMedeirosVarzielas:2006fc,deMedeirosVarzielas:2005qg,Altarelli:2005yx}.
In the relevant low energy
effective Yukawa operators the $SO(3)$ flavons
enter at the simplest possible one-flavon level, unlike $SU(3)$
where the lowest order operators must involve at least two flavons.

The existing analyses of models of this kind have so far been 
performed at the 4d effective non-renormalizable operator level
\cite{King:2005bj,deMedeirosVarzielas:2005ax,deMedeirosVarzielas:2006fc,deMedeirosVarzielas:2005qg}.
We have gone beyond existing analyses by considering,
as well as the 4d effective non-renormalizable operators,
also the underlying renormalizable 4d model in terms of a
high energy messenger sector. This represents an explicit ultraviolet
completion of the model. 
The messenger sector allows for effectively different
expansion parameters in the different charged sectors.
We performed a numerical analysis which 
shows that the model provides an excellent
fit to the charged fermion mass spectrum.
The model also predicts approximate tri-bimaximal
lepton mixing via CSD due to vacuum alignment
of flavon VEVs, with calculable deviations described by the neutrino
sum rule. The strong hierarchy in the charged
fermion sector, explained in terms of a small flavon VEV,
gets cancelled in the neutrino sector,
via the see-saw mechanism with sequential dominance,
leading to $m_2 \sim m_3$ for the lowest order neutrino masses, 
with the mild neutrino hierarchy $m_2 /m_3 \sim 1/5$  
produced by higher order corrections 
necessarily present in the model.

We have shown how the 
model can originate from a 5d orbifold GUT based on 
$SO(3)\times SO(10)$. From the persective of orbifold 
GUTs this provides significant progress since such models
have not so far provided a convincing explanation of 
fermion masses and mixings.
The small flavon VEVs responsible for the fermion mass hierarchy, 
which were postulated in an {\it ad hoc} way from
the 4d point of view, are seen to 
originate from bulk volume suppression in the 5d theory.
The bulk suppressed VEVs reduce the need for very high dimensional
operators, allowing a simpler operator structure which can be more
readily understood at the renormalizable level in terms
of an explicit messenger sector. 
The Pati-Salam symmetry is also shown to arise from a broken 
$SO(10)$ GUT. This demonstrates that, in the framework of a higher dimensional
theory, models based on the gauged Family Symmetry $SO(3)$,
or one of its discrete subgroups, can be fully consistent 
with $SO(10)$ Grand Unification.

To summarize, the model presented here provides a successful description
of fermion masses and mixings, with an excellent numerical
fit to the masses and mixings in the charged fermion sector, 
and a natural explanation of tri-bimaximal
lepton mixing. The framework of non-Abelian Family Symmetry
and GUTs, which has been used to account for tri-bimaximal
lepton mixing via CSD and vacuum alignment, has here been
combined with orbifold GUTs. The resulting synthesis allows
the fermion mass hierarchy to be explained by a small
bulk suppressed flavon VEV, which simplifies the Yukawa operator 
structure considerably, allowing the ultraviolet completion of the
model in terms of a renormalizable messenger sector.

%---------------------------

%---------------------------
\section*{Acknowledgement}
The authors are both extremely grateful to CERN where most of the
work in this paper was performed.
%---------------------------
%---------------------------
%\bibliographystyle{h-physrev}
%\bibliography{bibliography}

\end{document}